\documentclass{article}
\usepackage{amsmath}
\usepackage{graphicx}
\usepackage{amssymb}
\usepackage{subfigure}
\usepackage{cite}
\usepackage[margin=1in]{geometry}
\usepackage{epsfig}

\newcommand{\beq}{\begin{equation}}
\newcommand{\eeq}{\end{equation}}
\newcommand{\bea}{\begin{eqnarray}}
\newcommand{\eea}{\end{eqnarray}}

\newcommand{\gsim}{\lower.7ex\hbox{$\;\stackrel{\textstyle>}{\sim}\;$}}
\newcommand{\lsim}{\lower.7ex\hbox{$\;\stackrel{\textstyle<}{\sim}\;$}}

\newcommand{\bi}{\begin{itemize}}
\newcommand{\ei}{\end{itemize}}

\usepackage{color}
\usepackage{hyperref} 
\hypersetup{linktocpage=true}
\usepackage[all]{hypcap}
\hypersetup{
    bookmarks=true,         
    unicode=false,          
    pdftoolbar=true,        
    pdfmenubar=true,        
    pdffitwindow=false,     
    pdfstartview={FitH},    
    pdftitle={My title},    
    pdfauthor={Author},     
    pdfsubject={Subject},   
    pdfcreator={Creator},   
    pdfproducer={Producer}, 
    pdfkeywords={keyword1} {key2} {key3}, 
    pdfnewwindow=true,      
    colorlinks=true,       
    linkcolor=blue,          
    citecolor=magenta,        
    filecolor=magenta,      
    urlcolor=cyan           
}

\usepackage{fancyhdr}
\begin{document}
\thispagestyle{empty}
\vspace*{-22mm}
\begin{flushright}
UND-HEP-10-BIG\hspace*{.08em}07\\
TUM-HEP-791/11\\
\end{flushright}
\vspace*{10mm}

\vspace*{10mm}

\begin{center}
{\Large {\bf\boldmath
On $D\to X_u l^+ l^-$ within the Standard Model and Frameworks like the Littlest Higgs Model with T Parity}}
\vspace*{10mm}

{\bf Ayan\ Paul$^a$, Ikaros\,I.\ Bigi$^a$, Stefan\ Recksiegel$^b$} \\
\vspace{4mm}
{\small
$^a$ {\sl Department of Physics, University of Notre Dame du Lac}\\
{\sl Notre Dame, IN 46556, USA}\vspace{1mm}

$^b$ {\sl Physik Department, Technische Universit\"at M\"unchen,
D-85748 Garching, Germany}}


\vspace*{10mm}

{\bf Abstract}\vspace*{-1.5mm}\\
\end{center}
The $D\to X_u l^+ l^-$ transitions -- branching ratios, forward-backward asymmetry $A^c_{\rm FB}$,
the CP asymmetry $A^c_{\rm CP}$ and the CP asymmetry in the forward-backward asymmetry $A^{\rm CP}_{\rm FB}$ -- have two sources: for $D^{\pm}$ they represent a pure 
$\Delta C=1$ \& $\Delta Q=0$ current interaction whereas neutral $D$ mesons can also communicate via 
their anti-hadron. Standard Model (SM) contributions to $BR(D\to X_u l^+ l^-)$ come primarily from long distance dynamics, which overshadow short distance contributions by several orders of magnitude;  still they fall much below 
the present upper experimental bounds. Even the SM  contributions to $A^c_{\rm FB}$, $A^c_{\rm CP}$ and  $A^{\rm CP}_{\rm FB}$ are tiny, quite unlike in beauty hadrons. The branching ratios are hardly dented by contributions from 
the Littlest Higgs Models with T parity (LHT)
even in the short distance regime, let alone in the SM long distances dynamics. Yet the asymmetries $A^c_{\rm FB}$, $A^c_{\rm CP}$ and $A^{\rm CP}_{\rm FB}$ in these New Physics models can be enhanced over SM predictions, as they arise purely from short distance dynamics; this can occur 
in particular for $A^c_{\rm FB}$ and $A^{\rm CP}_{\rm FB}$ which get enhanced by orders of magnitudes. Even such enhancements hardly reach absolute sizes for observable experimental effects for $A^c_{\rm FB}$  and $A^c_{\rm CP}$.  
However LHT contributions to $A^{\rm CP}_{\rm FB}$ could be measured in experiments like the LHCb and the SuperB Collaboration. These results lead us to draw further conclusions on FCNCs within LHT-like models through some simple scaling arguments that encapsulate the essence of flavour dynamics in and beyond the Standard Model.

\noindent

\vspace*{10mm}
\newpage
\tableofcontents

\section{Introduction}
The discovery of charm quarks was and still is seen as a great success of the Standard Model (SM), since their
existence was necessary for the observed suppression of strangeness changing neutral currents. Charm hadrons were also found in
the expected mass range as was the predicted preference for decays to strange hadrons.

Yet, at last,  some possible hint of New Physics (NP) has appeared in charm physics. Compelling evidence 
for $D^0-\bar D^0$ oscillations has been presented by Belle, BaBar and CDF \cite{D0obs}. The HFAG has 
combined the results on neutral $D$ decays allowing CP violation \cite{HFAGCHARM}\footnote{Up to date results can be found in the \href{http://www.slac.stanford.edu/xorg/hfag/charm/CHARM10/results_mix+cpv.html}{HFAG} website}:
\bea
\nonumber x_D = \frac{\Delta M_D}{\Gamma_D} = 0.63 ^{+0.19}_{-0.20} \; &,& \; 
y_D = \frac{\Delta \Gamma_D}{2\Gamma_D} = 0.75 \pm 0.12 \\
\left |\frac{q}{p}  \right | = 0.91 ^{+0.18}_{-0.16} \; &,& \; 
\phi   = - 10.2 ^{+9.4}_{-8.9}\; (^o)
\label{DOSCDATA}
\eea 
 The observation of $D^0-\bar D^0$ oscillations is hardly disputed, while the relative size of $x_D$ and 
 $y_D$ is not clear yet. Before these experimental results, most theorists argued that the 
 SM predicts $x_D$, $y_D$ $\leq 10^{-4}$ -- yet not all: in 1998, $x_D$, $y_D$ $\leq 10^{-2}$ was  called a 
 SM {\em conservative} bound \cite{VAR98}; in 2000 and 2003 a SM prediction obtained from a 
 sophisticated operator product expansion yielded $x_D$, $y_D$ $\sim {\cal O}(10^{-3})$ \cite{DUAL} and more recently in \cite{LENZ}; 
 alternatively, 
 in 2001 and 2004 a SM prediction on $D^0-\bar D^0$ oscillations was based on $SU(3)$ breaking mostly in the phase space for $y_D$ and then from a dispersion relation for $x_D$ \cite{FALK}.
 
While the present experimental results on $x_D$ and $y_D$ can be accommodated within the available theoretical SM estimates, and no non-zero CP asymmetry has been seen yet, the observation of $D^0 - \bar D^0$ oscillation, however, has `wetted' 
the appetite on thinking of NP in charm decays. The authors of Ref.\cite{NIRETAL} consider a (approximately) 
$SU(2)_L$-invariant NP scenario; therefore NP contributions to $D^0 - \bar D^0$ and $K^0 - \bar K^0$ oscillations are not 
independent of each other. 

There is a large variety of NP models, in which the Higgs boson appears as a pseudo-Nambu-Goldstone boson of a 
spontaneously broken global symmetry, namely the `Little Higgs' class of models \cite{LH,SimHiggs}\footnote{For an overview of the different `flavours' of Little Higgs models cf. \cite{Perel}}.  To achieve this 
program one needs at least heavy gauge bosons $W_H^{\pm}$, $Z_H$ and $A_H$, a heavy top partner $T$ 
and a scalar triplet $\Phi$ as physical degrees of freedom as is implemented in the `Littlest Higgs' model \cite{L2H,Han}. Studying electroweak precision observables shows that for such new states to arise below the 
1 TeV scale, one needs an additional discrete symmetry \cite{CL1,CL2}, called T parity: the SM particles and the heavy 
top partner $T$ are {\em even} and $W_H^{\pm}$, $Z_H$, $A_H$ and $\Phi$ are {\em odd}. A consistent 
implementation of T parity requires also the introduction of the so-called `mirror' fermions -- one for each quark and 
lepton species -- that are {\em odd} under T parity \cite{CL2,Low}. This creates the `Littlest Higgs' model with 
T parity (LHT)\footnote{For a detailed description of the Littlest Higgs Model with T parity cf. \cite{LHTRev}.}. While 
{\em some} theorists probably see it as intellectually economical, most experimentalists do not 
view it like that; at the same time they should understand that it can provide them with more work, but less 
so than SUSY! The most important point here is that the motivation for LHT models come from {\em outside} flavour 
dynamics; at the same time they can create important non-trivial signals of NP in $B$, $K$ -- and $D$ physics. 
The LHT models can -- not necessarily, but possibly -- affect $\Delta C =2$ dynamics significantly. In particularly, 
it can generate sizable or even relatively large 
indirect CP violation in $D^0$ decays \cite{DKdual}. It can implement a {\em dynamical} realization of the symmetry 
approach described in Ref.\cite{NIRETAL}.  

Encouraged by the findings of Ref. \cite{DKdual} about the possible impacts of LHT models in $\Delta C =2$ dynamics we had looked at two $\Delta C=1$ processes, 
$D^0 \to \gamma\gamma$ and $D^0\to \mu^+\mu^-$ in a previous study \cite{PBR1}. 
 While LHT failed to contribute significantly to the total decay rates in these channels, it had led us to some general conjectures on charm changing neutral currents 
(CCNC) within LHT-like scenarios. In this current work we continue to analyze the impact of LHT on another 
$\Delta C =1$ process to probe deeper into CCNCs in LHT-like scenarios, namely to $D\to X_ul^+l^-$. Similar to our previous work, we do not see any sizable enhancements in the global decay rates. However in the presence of 
{\em large} weak phases in LHT-like models it might seem surprising to find also very tiny contributions to the 
CP and forward-backward asymmetries $A_{\rm CP}^c$ and $A^c_{\rm FB}$. This is because the (non-trivial) SM asymmetries get produced by {\em short} distance dynamics, and 
LHT-like scenarios can create much larger $A_{\rm CP}^c$ and $A^c_{\rm FB}$ than the SM can; yet they are still 
small in their absolute size. However the LHT contributions to the forward-backward CP asymmetry $A^{\rm CP}_{\rm FB}$ 
are sizable -- even large -- such that they can be experimentally measured in the coming decade.  

In this article, we discuss both short and long distance SM contributions to $D\to X_u l^+l^-$ in Sect.\ref{SMPRED} along with $A_{\rm CP}^c$,  $A^c_{\rm FB}$ and $A^{\rm CP}_{\rm FB}$ in this channel. We will describe mostly 
$D^\pm\to X_u l^+l^-$, since it 
is given only by $\Delta C = 1$ couplings, while one can also produce $D^0\to X_u l^+l^-$ by $\Delta C = 2$ couplings due to $D^0 - \bar D^0$ oscillations. We go on to briefly introduce LHT and its contributions to 
$D\to X_u l^+l^-$ in Sect.\ref{LHT}. Our quantitative findings are presented in Sect.\ref{RESULTS}. We take a more critical look at CCNCs in LHT-like models in Sect.\ref{FCNC}. In Sect.\ref{BOXPEN} we put forward some simple scaling arguments to explain why NP interventions such as in LHT-like models have effects of the size we see and the conclusion to this work follows in Sect.\ref{CON}.

\boldmath
\section{SM Contributions to $D\to X_u l^+ l^-$}
\unboldmath
\label{SMPRED}

The transition of $D\to X_u l^+l^-$ must be produced by charm changing neutral currents, 
which are much weaker even than their strangeness and beauty analogues in the SM. 
These decay rates are 
tiny and dominated by long distances effects. 
Yet the forward-backward $A^c_{\rm FB}$, CP asymmetries $A^c_{\rm CP}$  and CP asymmetry in the forward-backward asymmetry $A^{\rm CP}_{\rm FB}$ could still be controlled by SM {\em short} distances dynamics. However the dynamics scenery 
is very complex as shown below; conceptually very similar to $B\to X_s l^+l^-$, but quantitatively at a smaller 
level. One can learn a lot  by studying these {\em asymmetries} about SD dynamics in general where 
different operators mix and an alternative perspective can be obtained from $B$ studies. It turns out -- 
not surprisingly -- that a careful scrutiny needs huge statistics. This would be an important task for a 
Super-B factory like the recently approved SuperB project undertaken by the INFN \cite{SuperB}; LHCb might also be able to address it. 

We will discuss first $D^{\pm} \to X_u l^+l^-$, since it proceeds purely by a $\Delta C=1$ interaction; then we 
will comments on lessons learnt from neutral $D \to X_u l^+l^-$ transitions, where $D^0 - \bar D^0$ oscillations 
can get involved.

\boldmath
\subsection{$\Gamma_{\rm SM}(D\to X_u l^+l^-)$}
\unboldmath
\label{DECAY}

The quark level process $c \to u l^+l^-$ is described with an operator basis of the following ten operators: 

\begin{eqnarray}
\nonumber&&O_1^{(q)}=(\bar{u}_L^\alpha\gamma_\mu q_L^\beta)(\bar{q}_L^\beta\gamma^\mu c_L^\alpha)\hspace{1.2cm}O_2^{(q)}=(\bar{u}_L^\alpha\gamma_\mu q_L^\alpha)(\bar{q}_L^\beta\gamma^\mu c_L^\beta)\label{O12}\\
\nonumber&&O_3=(\bar{u}_L^\alpha\gamma_\mu c_L^\alpha)\sum_q(\bar{q}_L^\beta\gamma^\mu q_L^\beta)\hspace{0.85cm}O_4=(\bar{u}_L^\alpha\gamma_\mu c_L^\beta)\sum_q(\bar{q}_L^\beta\gamma^\mu q_L^\alpha)\label{O34}\\
\nonumber&&O_5=(\bar{u}_L^\alpha\gamma_\mu c_L^\alpha)\sum_q(\bar{q}_R^\beta\gamma^\mu q_R^\beta)\hspace{0.85cm}O_6=(\bar{u}_L^\alpha\gamma_\mu c_L^\beta)\sum_q(\bar{q}_R^\beta\gamma^\mu q_R^\alpha)\label{O56}\\
\nonumber&&O_7=\frac{e}{16\pi^2}m_c(\bar{u}_L\sigma_{\mu\nu}c_R)F^{\mu\nu}\hspace{0.75cm}O_8=\frac{g_s}{16\pi^2}m_c(\bar{u}_L\sigma_{\mu\nu}T^ac_R)G^{\mu\nu}_a\label{O78}\\
&&O_9=\frac{e^2}{16\pi^2}(\bar{u}_L\gamma_\mu c_L)(\bar{l}\gamma^\mu l)\hspace{1.1cm}O_{10}=\frac{e^2}{16\pi^2}(\bar{u}_L\gamma_\mu c_L)(\bar{l}\gamma^\mu\gamma_5 l)\label{O910}
\end{eqnarray}
where $q=d,s,b$ and $\alpha,\beta$ are colour indices. The charm operators $O_i$, $i=1, ..., 10$ are 
analogous to those in $b$ decays \cite{Burd95}. The effective weak Hamiltonian is expressed in terms of 
these operators taken at a scale $\mu$: 
\beq
\mathcal{H}_{eff}(\mu )= 
-4\frac{G_F}{\sqrt{2}}\left[\sum_{q}C_1^{(q)}(\mu)O_1^{(q)}(\mu)+C_2^{(q)}(\mu)O_2^{(q)}(\mu)+
\sum_{i=3}^{10}C_i(\mu)O_i(\mu)\right]\\
\eeq
where the coefficients describe renormalization of the operators at $\mu$ from the normalization scale. 
For the starting point one can naturally chose $M_W$:  
\begin{eqnarray}
\nonumber&&C^{(q)}_1(M_W)=0\phantom{xx}C^{(q)}_2(M_W)=1\phantom{xx}C_{3-6}(M_W)=0 \\
\nonumber&&C_7(M_W)=-\frac{1}{2}\sum_{j=d,s}V^*_{uj}V_{cj}\left(D^\prime_0(x_j)-D^\prime_0(x_b)\right) \\
\nonumber&&C_8(M_W)=-\frac{1}{2}\sum_{j=d,s}V^*_{uj}V_{cj}\left(E^\prime_0(x_j)-E^\prime_0(x_b)\right) \\
\nonumber&&C_9(M_W)=\sum_{j=d,s}V^*_{uj}V_{cj}\left(\frac{Y_0(x_j)}{\sin^2(\theta_W)}-4C_0(x_j)-D_0(x_j)-(x_j\rightarrow x_b)\right) \\
&&C_{10}(M_W)=\sum_{j=d,s}V^*_{uj}V_{cj}\left(\frac{Y_0(x_j)-Y_0(x_b)}{\sin^2(\theta_W)}\right) \; , 
\end{eqnarray}
where $x_j=m_j^2/m_W^2$, with $m_j$ being the masses of the internal down type quarks $j=d,s,b$. Here we have used the unitarity of the CKM matrix, 
$\sum_{j=d,s,b}V^*_{uj}V_{jc}=0$ to eliminate the dependence of the Wilson coefficients on the third family. The limit $x_d\to 0$ cannot be taken as there is a logarithmic divergence in $D^0(x)$ in that limit. 
The form factors are defined in a modified way from 
Ref.\cite{InamiLim}:  

\begin{eqnarray}
\nonumber C_0(x)&=&\frac{1}{2}\left(\frac{x}{4}-\frac{3}{8}\frac{1}{(x-1)}+\frac{3}{8}\frac{2x^2-x}{(x-1)^2}\log(x)+\gamma(x)\right)\\
\nonumber D_0(x)&=&Q_d E_0(x)+\frac{x \left(9-43 x+28 x^2\right)}{12 (1-x)^3}+\frac{x \left(21-66 x+41 x^2-2 x^3\right)}{12 (1-x)^4}\log(x)-2\gamma(x)\\
\nonumber E_0(x)&=&-\frac{2}{3}\log(x)+\frac{x^2\left(15-16x+4x^2\right)}{6(1-x)^4}\log(x)+\frac{x\left(18-11x-x^2\right)}{12(1-x)^3}
\end{eqnarray}
\begin{eqnarray}
\nonumber D^\prime_0(x)&=&Q_dE^\prime_0(x)-\left(\frac{1}{3}+\frac{11x^2-7x+2}{4(1-x)^3}+\frac{6x^3}{4(1-x)^4}\log(x)\right)\\
\nonumber E^\prime_0(x)&=&-\frac{5}{12}+\frac{1-5x-2x^2}{4(x-1)^3}+\frac{3x^2}{2(x-1)^4}\log(x)\\
Y_0(x)&=&\frac{x}{8}\left(\frac{x-4}{x-1}+\frac{3x}{(x-1)^2}\log(x)\right)\label{eq:SM}
\end{eqnarray}
$Q_d$ is the charge of the internal down type quarks. Here $\gamma(x)$ is the gauge dependent term which, for $\xi=1$, is \footnote{Although some of the formfactors have gauge dependence, it drops out of the final amplitude as expected.}
\begin{eqnarray}
 \gamma(x)=\frac{7}{8}\left(\frac{x}{(x-1)^2} \log(x)-\frac{1}{x-1}\right)
\end{eqnarray}
At scales $\mu < M_W$ one can express $\mathcal{H}_{eff}(\mu )$ using $C_i(\mu )$, $i= 1, ..., 10$ evolving 
through the two-loop QCD renormalization group equation. Actually one has two regimes, $M_W$ to $m_b$ 
and $m_b$ to $m_c$\footnote{Of course, one can question the robustness of the selection of $\mu=m_b$ and $\mu=m_c$. However, any arguments on this choice are purely acedemic at this point considering the level of precision aimed at in the current calculations.}; a matching condition as usual is applied through 
$\alpha_s (m_b,m_b; 4) = \alpha_s (m_b,m_b; 5)$. These operators mix via renormalization; in 
particular $O_7$ mixes with $O_1$, $O_2$ \cite{Burd95,Greub}, $O_{3 - 6}$ and $O_8$. However, as pointed out in \cite{Greub}, $O_7$ is completely dominated by the two loop QCD radiative correction, which was taken into account in \cite{Burd,Fajfer3}: 
\begin{eqnarray}
\nonumber C_7(m_c)&=&\eta_c^{\frac{16}{25}}\eta_b^{\frac{16}{23}}C_7(m_W)-\frac{16}{3}\left(\eta_c^{\frac{14}{25}}\eta_b^{\frac{14}{23}}-\eta_c^{\frac{16}{25}}\eta_b^{\frac{16}{23}}\right)C_8(m_W)-V^*_{ub}V_{cb}\sum _{i=1}^8\sum _{j=1}^6 C_j(m_b)X_{ji} \eta_c^{z_i}\\
\nonumber&&+\frac{\alpha_s(m_c)}{4\pi }C_2(m_c)\left[V^*_{us}V_{cs}\left\{f^2\left(\frac{m_s}{m_c}\right)-f^2\left(\frac{m_d}{m_c}\right)\right\}+V^*_{ub}V_{cb}f^2\left(\frac{m_d}{m_c}\right)\right]\\
\end{eqnarray}
where
\begin{eqnarray}
\nonumber \eta_b&=&\frac{\alpha_s(m_W)}{\alpha_s(m_b)},\phantom{x}\eta_c=\frac{\alpha_s(m_b)}{\alpha_s(m_c)}\\
\end{eqnarray}
The matrix ${\bf X}$, the vector ${\bf z}$ and the function $f$ introduced above are given in Appendix \hyperref[APPA]{A} along with the Wilson coefficients for the operators $O_{1 - 6}$ both at $\mu=m_b$ and $\mu=m_c$. We include the next to leading order correction to the running of the $\alpha_s$  with $\alpha_s(m_W)=0.125$, $\alpha_s(m_b)=\alpha_s(m_b,m_W,5)$, $\alpha_s(m_c)=\alpha_s(m_c,m_b,4)$.
\begin{eqnarray}
\nonumber\alpha_s(\mu,\mu^\prime,n_f)&=&\frac{\alpha_s(\mu^\prime,\mu^\prime,n_f)}{v(\mu,\mu^\prime,n_f)}\left(1-\frac{\beta_1(n_f)}{\beta_0(n_f)}\frac{\alpha_s(\mu^\prime,\mu^\prime,n_f)}{4\pi}\frac{\log(v(\mu,\mu^\prime,n_f))}{v(\mu,\mu^\prime,n_f)}\right)\\
\nonumber  v(\mu,\mu^\prime,n_f)&=&1-\beta_0(n_f)\frac{\alpha_s(\mu^\prime,\mu^\prime,n_f)}{2\pi}\log(\frac{\mu^\prime}{\mu})\\
\nonumber\beta_0(n_f)&=&11-\frac{2}{3}n_f\\
\beta_1(n_f)&=&102-\frac{38}{3}n_f
\end{eqnarray}
$O_9$ mixes with $O_{1-6}$ beyond the leading order. However, only $C_1$ and $C_2$ are numerically significant and lead to an important cancellation amongst themselves. As we shall later see, the dominant contribution to the SM branching fraction comes from $O_9$; hence it is important to have a detailed look at it contrary to what was argued in \cite{Burd,Fajfer3}. The Wilson coefficient for $O_9$ after QCD corrections is given by 
\footnote{The dependence of $C_9(m_c)$ on $h(1,s)$ and $h(0,s)$ have been ignored as they numerically and conceptually insignificant in this case.}
\begin{eqnarray}
C_9(m_c)=C_9(m_W)+\sum_{j=d,s}V^*_{uj}V_{cj}h\left(z_j,\hat{s}\right)\left(\sum_{i=1,3,5}3\bar{C}_i(m_c)+\bar{C}_{i+1}(m_c)\right)
\label{eq:witherr}
\end{eqnarray}
where $\bar{C}_i$ are defined in~\cite{Beneke2001,deBoer2016}\footnote{The authors of \cite{deBoer2016} pointed out an error in Eq.\ref{eq:witherr} leading to the incorrect definition of $\bar{C}_i$. We have checked our results after fixing the error and the numerical effects are very minor, barely modifying most of the plots. The conclusions of our paper remain unchanged.}. The function $h(z_j,s)$ comes from the one loop QCD correction to the four fermion operator and is given by 
\footnote{The overall sign of $h(z,\hat{s})$ is incorrect in \cite{Burd}.}
\begin{eqnarray}
\nonumber h(z,\hat{s})&=&Q_d \tilde{h}(z,\hat{s})\\
\nonumber \tilde{h}(z,\hat{s})&=&-\frac{4}{3}\log\frac{m_c}{\mu}-\frac{4}{3}\log(z)+\frac{4}{9}+\frac{8}{3}\frac{z^2}{s}-\frac{1}{3}\left(2+\frac{4z^2}{s}\right)\sqrt{\left|1-\frac{4z^2}{s}\right|}
\times\left\{
\begin{array}{ccc}
2\tan^{-1}\frac{1}{\sqrt{\frac{4z^2}{s}-1}} & &\mbox{if $\hat{s} < 4z^2$}\\
\log\left|\frac{\sqrt{1-\frac{4z^2}{s}}+1}{\sqrt{1-\frac{4z^2}{s}}-1}\right|-i \pi & &\mbox{if $\hat{s} > 4z^2$}\\
\end{array}
\right.\\
\end{eqnarray}
with
\begin{eqnarray}
\nonumber \hat{s}=\frac{(p_{l^+}+p_{l^-})^2}{m_c^2},\phantom{xxx}z_j=\frac{m_j}{m_c}
\end{eqnarray}
As in the case of the analogous decay in $B$ systems, the logarithmic term in $h(z,\hat{s})$ exactly cancels the logarithmic dependence in $C_9(m_W)$ that comes from the electromagnetic penguin and hence removes the logarithmic dependence on light quark masses at the scale $\mu=m_W$ as was pointed out in \cite{Burd}. The importance of QCD correction was pointed out in \cite{Fajfer4}, but we disagree with their argument that the purely electroweak ``Inami-Lim'' contribution to $C_9$ should be ignored as it is dependent on light quark mass and is reproduced as a limit of the QCD correction when $\hat{s}\to 0$.  A careful look at the form of $h(z,\hat{s})$ shows that the logarithms have opposite signs in the ``Inami-Lim'' term and in the QCD correction as argued before\footnote{The incorrect argument in \cite{Fajfer4} stems from an incorrect relative sign between the ``Inami-Lim'' term and the QCD correction.}. A discussion of the logarithmic dependence of $C_9$  before including QCD corrections  and its cancellation after including the same is discussed in \cite{Grin} for the case of $B$ mesons. A similar argument applies in the case of the $D$ mesons too.

$O_{10}$ does not suffer from any QCD corrections \cite{Burd, Buras95} which makes the assumption made in 
\cite{Fajfer3} unnecessary \footnote{$C_{10}$ is indeed very tiny in the SM as we shall show and is also stated in 
\cite{Fajfer4}. Hence, $C_{10}(m_c)=C_{10}(m_W)$. However we keep this contribution as it is important in 
$A^c_{\rm FB}$ and has potentials of being largely enhanced by LHT as we saw in \cite{PBR1}.}. Finally, the differential decay branching fraction is given by
\begin{eqnarray}
\nonumber &&\frac{d}{d\hat{s}}Br_{\rm SM}^{\rm SD}\left(D\to X_u l^+l^-\right)=\\\nonumber&&\frac{1}{\Gamma _D}\frac{G_F^2\alpha ^2m_c^5}{768\pi^5}(1-\hat{s})^2
\Bigg[\left(\left|C_9(\mu)\right|^2+\left|C_{10}(\mu)\right|^2\right)(1+2\hat{s})
+12 \text{ Re}(C_7(\mu)C^*_9(\mu))+4\left(1+\frac{2}{\hat{s}}\right)\left|C_7(\mu)\right|^2\Bigg]\\
\end{eqnarray}
with $\mu=m_c=1.2$ GeV. Integrating over $\hat{s}$ gives us the total decay rate. One has to be careful about not picking up the infrared divergence in the differential decay rate. We made an infrared cut on $\hat{s}$ at about an invariant dilepton momentum of $20$ MeV. We get a branching fraction of
\begin{eqnarray}
BR_{\rm SD}^{\rm SM}(D \to X_u e^+e^-)\sim 3.7 \times 10^{-9}
\end{eqnarray}
 which is smaller than what is stated in \cite{Burd,Fajfer3,Fajfer2} but larger than the number in \cite{Fajfer4} for reasons stated above. The same for muons in the final state is slightly smaller due to the finite mass of the muon. The SD SM contribution is dominated by $C_9$ primarily with contributions from the purely electroweak part, coming almost entirely from the electromagnetic penguin, and  an order of magnitude smaller contribution from the QCD correction coming from the four fermion operators $O_1$ and $O_2$. Contrary to what is stated in \cite{Fajfer4}, $C_7$ provides only a subdominant contribution in spite of its huge enhancement from the two loop $O(\alpha_s)$ contributions.
 
 As we stated before, the SD contribution in SM is completely overshadowed by the LD contribution that comes from intermediate vector meson states. These resonance contributions lead to a branching fraction estimated in 
 \cite{Burd}: 
 \beq
BR_{\rm LD}^{\rm SM}(D \to X_u e^+e^-) = BR^{\rm SM}(D \to X_u e^+e^-)\sim {\cal O}(10^{-6}) 
\label{DPMBR}
 \eeq

Rough estimates can show that the SD contributions for the branching ratio are much smaller than 
LD contributions. Why did we (and other authors) undertake a time consuming OPE analysis? 
Finding incorrect statements in published literature maybe be intellectually acceptable, even if 
such statements are of only academic significance. However we are driven by matters of much more practical interest: 
Some asymmetries on which distributions are based give a more direct access to SD dynamics, namely
$A^c_{\rm FB}$ and $A^c_{\rm CP}$ as we shall discuss next. 

\boldmath
\subsection{$A^c_{\rm FB}$ and $A^c_{\rm CP}$ and $A^{\rm CP}_{\rm FB}$}
\unboldmath
\label{ASYMM}

Asymmetries between $D^{\pm} \to X_u l^+l^-$ have a single source, namely $\Delta C=1$, $\Delta Q=1$ 
currents. Later we will comment on $D^0 \to X_u l^+l^-$ vs. $\bar D^0 \to X_u l^+l^-$, where $D^0 - \bar D^0$ 
oscillations, in principle, could get involved. 

The normalized forward-backward asymmetry is defined from the double differential decay rate as
\begin{eqnarray}
A^c_{\rm FB}(\hat{s})=\frac{\int_{-1}^{1}\left[\frac{d^2}{d\hat{s}dz}\Gamma(D^+\to X_ul^+l^-)
- \frac{d^2}{d\hat{s}dz}\Gamma(D^-\to X_ul^+l^-) \right]{ \rm sgn}(z)dz}
{\int_{-1}^{1}\left[d^2\Gamma(D^{\pm}\to X_ul^+l^-)/d\hat{s}dz\right]dz}
\end{eqnarray}
After performing the integral over the angular distribution we get
\begin{eqnarray}
\nonumber &&A^c_{\rm FB}(\hat{s})=\frac{-3\left[ \Re (C_{10}^*(\mu) C_9(\mu))\hat{s} +2 \Re (C_{10}^*(\mu) C_7(\mu)\right]}{(1+2s)\left(\left|C_9(\mu)\right|^2+\left|C_{10}(\mu)\right|^2\right)+4\left|C_7(\mu)\right|^2\left(1+\frac{2}{s}\right)+12\Re \left(C_7(\mu) C_9^*(\mu)\right)}\\
\end{eqnarray}
Since $C_{10}$ is real, $A^c_{\rm FB}(\hat{s})$ picks up the real part of $C_9$ and $C_7$ which are both in general complex. Integrating over $\hat{s}$ we get
\begin{eqnarray}
A^c_{\rm FB}\sim 2\times 10^{-6}
\end{eqnarray}
$A^c_{\rm FB}(\hat{s})$ is mostly proportional to $C_{10}$ which is tiny in the SM as expected from the suppression of FCNC in charm physics. Unlike the integrated  decay rate, $A^c_{\rm FB}$ is not very sensitive to infrared divergences. Since the angular distribution of the double differential decay rate is almost opaque to the LD contribution from SM\footnote{A detailed argument on this can be found in \cite{AliMan} for the $B$ mesons. A similar argument holds for the $D$ mesons too.}, $A^c_{\rm FB}$ absorbs purely SM SD contributions -- yet truly tiny! 

The CP asymmetry parameter $A^c_{\rm CP}(\hat{s})$ is defined as
\begin{eqnarray}
A^c_{\rm CP}(\hat{s})=\frac{\frac{d}{d\hat{s}}\Gamma(D^+\to X_u l^+l^-)-\frac{d}{d\hat{s}}\Gamma(D^-\to X_{\bar{u}} l^+l^-)}{\frac{d}{d\hat{s}}\Gamma(D^+\to X_u l^+l^-)+\frac{d}{d\hat{s}}\Gamma(D^-\to X_{\bar{u}} l^+l^-)}
\end{eqnarray}
In general any Wilson coefficient $C_i(\mu)$ in the differential decay rate can be written as 
\begin{eqnarray}
\nonumber C_i(\mu)=\xi^0_i+\lambda^j_i\xi^j
\end{eqnarray}
where summation over $j$  is implied. For $\Gamma(\bar{D}\to X_{\bar{u}} l^+l^-)$, $\lambda_j\to \lambda^*_j$. The numerator will have contributions of the type
\begin{eqnarray}
\nonumber \left|C_i(\mu)\right|^2&\to&\Im(\lambda^l\lambda^{m*})\Im(\xi_i^l\xi_i^{m*})+2 \Im (\lambda^l)\Im(\xi_i^0\xi_i^{l*})\\
\nonumber \Re\left(C_i(\mu)C^*_j(\mu)\right)&\to&\Im(\lambda^l\lambda^{m*})\Im(\xi_i^l\xi_j^{m*})+\Im(\lambda^{l})\Im(\xi_i^{l}\xi_j^{0*})+\Im(\lambda^{m*})\Im(\xi_i^{0}\xi_j^{m*})\label{eq:num}
\end{eqnarray}
The denominator will have contributions of the type
\begin{eqnarray}
\nonumber \left|C_i(\mu)\right|^2&\to&\Re(\lambda^l\lambda^{m*})\Re(\xi_i^l\xi_i^{m*})+2 \Re (\lambda^l)\Re(\xi_i^0\xi_i^{l*})+\left|\xi_i^0\right|^2\\
 \Re\left(C_i(\mu)C^*_j(\mu)\right)&\to&\Re(\lambda^l\lambda^{m*})\Re(\xi_i^l\xi_j^{m*})+\Re(\lambda^{l})\Re(\xi_i^{l}\xi_j^{0*})+\Re(\lambda^{m*})\Re(\xi_i^{0}\xi_j^{m*})+\Re(\xi_i^0\xi_j^{0*})\label{eq:den}
\end{eqnarray}
In the limit that only $C_9$ has an imaginary component and terms proportional to $V^*_{ub}V^{}_{cb}$ are ignored, the results stated in \cite{Krug} are realized. The numerator is sensitive only to the imaginary contributions from the Wilson coefficients to the decay rate and the denominator to the real contributions as should be the case;  for 
the relative phases in the matrix element is needed for a CP asymmetry. Unlike in the $B$ mesons, in the case of 
$D\to X_u l^+l^-$ only $C_{10}$ is purely real, and hence both $C_7$ and $C_9$ contribute to $A_{\rm CP}^c$. The integrated asymmetry $A^c_{\rm CP}$ in SM turns out to be
\begin{eqnarray}
A^c_{\rm CP}=\frac{\Gamma(D^+\to X_u l^+l^-)-\Gamma(D^-\to X_{\bar{u}} l^+l^-)}
{\Gamma(D^+\to X_u l^+l^-)+\Gamma(D^-\to X_{\bar{u}} l^+l^-)}\sim3\times 10^{-4} \; ; 
\end{eqnarray}
i.e., still tiny. 
 We also find that the bulk of $A^c_{\rm CP}$ comes from $C_7$ and is due to the presence of the two loop $O(\alpha_s)$ contribution. Indeed, the $C_9$  contribution, which comes from  the mixing of $O_{1-6}$ with $O_{9}$, serves only to suppress this contribution by an order of magnitude all of which stand in stark contrast to what happens in the analogous decay of the $B$ mesons. Both of these contributions arise only after the inclusion of QCD radiative corrections and both of which are proportional to $V^*_{us}V^{}_{cs}$; ignoring the term proportional to 
$V^*_{ub}V^{}_{cb}$ as it is relatively much smaller.

 We can also look at $A^{\rm CP}_{\rm FB}(\hat{s})$ which is the normalized difference in the forward-backward asymmetry in $D\to X_u l^+l^-$ and $\bar{D}\to X_{\bar{u}} l^+l^-$ defined as \cite{BHI}
\begin{eqnarray}
A^{\rm CP}_{\rm FB}(\hat{s})=\frac{A^c_{\rm FB}(\hat{s})+A^{\bar{c}}_{\rm FB}(\hat{s})}{A^c_{\rm FB}(\hat{s})-A^{\bar{c}}_{\rm FB}(\hat{s})}
\end{eqnarray}
In the limit of CP symmetry $A^c_{\rm FB}(\hat{s})$ and $A^{\bar{c}}_{\rm FB}(\hat{s})$ have to be exactly equal in magnitude but with an opposite sign \cite{Krug,SRC}. As the forward-backward asymmetry is defined in terms of the positive 
anti-lepton, $A^c_{\rm FB}(\hat{s})$ and $A^{\bar{c}}_{\rm FB}(\hat{s})$ have opposite signs. $A^{\rm CP}_{\rm FB}(\hat{s})$ is sensitive to the phase in $C_7$,  $C_9$ and $C_{10}$. The SM offers phases only in $C_7$ and $C_9$ in 
$D\to X_u l^+l^-$ and none in $C_{10}$. Hence the integrated asymmetry turns out to be tiny.
\begin{equation}
\int A^{\rm CP}_{\rm FB}(\hat{s}) d\hat{s}=A^{\rm CP}_{\rm FB}\sim3\times 10^{-5}
\end{equation}
If NP brings about any new phases in either $C_7$,  $C_9$ and $C_{10}$, $A^{\rm CP}_{\rm FB}$ stands a chance of large enhancements.

Putting together $A^c_{\rm FB}$, $A^c_{\rm CP}$  and $A^{\rm CP}_{\rm FB}$ gives us a good insight into the sizes of the phases in the Wilson coefficients.  $A^c_{\rm FB}$ is sensitive to the size of $C_{10}$ and the real parts of $C_7$ and $C_9$, while $A^c_{\rm CP}$ gives us an idea of the size of the phases in $C_7$ and $C_9$ and $A^{\rm CP}_{\rm FB}$ is sensitive to phases in all $C_7$,  $C_9$ and $C_{10}$. Within the SM, we can conclude from our numbers, the size of $C_{10}$ is extremely small, as it should be, since it suffers from a very strong GIM suppression, and it also lacks a phase. Both $C_7$ and $C_9$ have phases because of the QCD corrections which are more prominent in the case of $D$ mesons than in $B$ mesons as the purely electroweak contribution is truly tiny. Along with the decay rate, these provide us a very useful tool to probe into the flavour structure of any NP models and new sources of FCNC.

\subsection{A Note on LD Dynamics' Impact  on the Asymmetries}
\label{ASYMMLD}

The asymmetries discussed in the previous section do not incorporate any SM LD contributions, neither in their extraction from the difference in the partial decay rates nor in their normalization. This might be unreasonable 
considering that SM SD contributions yield a branching ratio of only $1.5 \times 10^{-9}$, while SM LD yields something 
like $O(10^{-6})$ even considering that LD estimates come with very large uncertainties. 
Let us add a comment why 
LD physics have only a little impact on these asymmetries, either from the difference in the numerator or the normalization in 
the denominator. 

For $A^c_{\rm CP}$ and $A^{\rm CP}_{\rm FB}$ it is obvious that SM LD physics cannot contribute a CP violating phase, and any CP asymmetry in this process has to come from SD physics whatever the origin is -- SM or NP.  
$A^c_{\rm FB}$ is more sensitive to SM LD `pollution' in its definition as the difference in the hemispherical integral of the double differential decay rate. SM LD contributions to it will primarily come through final state interactions and the dominance of light internal quarks in this process and quantitative statements on their sizes require separate analyses on each exclusive process. SM LD physics can also make their presence felt in all these observables through the normalizations 
entering the definitions of these asymmetries. A natural way to remove such LD contributions is to cut off 
the dileptonic mass distribution around the $\rho$, $\omega$ and $\phi$ widths. 

Making such cuts we find that it will {\em decrease} $A^c_{\rm FB}$ 
and {\em increase} $A^c_{\rm CP}$ and $A^{\rm CP}_{\rm FB}$ by around $10\%$ to $20\%$. 
Not surprisingly, these cuts affect $A^c_{\rm FB}$ more than $A^c_{\rm CP}$ and $A^{\rm CP}_{\rm FB}$. 
Similar effects arise, when we include LHT contributions as discussed below. More sophisticated cuts could yield even better 
results. As pointed out below, only $A^{\rm CP}_{\rm FB}$ has a realistic chance to be measurable by LHCb and a 
Super-Flavour Factory. It seems to us that such a theoretical uncertainty is fully acceptable for a 
search for NP as of now. 

As a final point: including non-resonant LD effects and making a cut on the resonances 
seem to produce opposing effects and can well nullify each other; however only a detailed study can resolve 
this issue.

Throughout the rest of the article we will stick to the definition of the asymmetries in terms of SD operators and infer on our results accordingly keeping in mind that a sufficiently motivated reader will already have had gone through this section by then. Any NP contribution to these asymmetries  are strictly SD.

\boldmath
\subsection{Comments on $D^0/\bar D^0 \to X_u l^+l^-$}
\unboldmath
\label{D0DECAY}

The branching ratios for neutral $D\to  X_u l^+l^-$ are again dominated by SM LD contributions and their size is comparable to what is stated in Eq.\ref{DPMBR}, 
namely of order $10^{-6}$. For $A^c_{\rm FB}$ and $A^c_{\rm CP}$ one has a much more complex system in hand for neutral 
$D$ mesons, because $D^0 - \bar D^0$ oscillations have been found on the level of $0.5\% - 1$ \% for $x_D$ and $y_D$, 
see Eq.\ref{DOSCDATA}, 
which might be still consistent with the SM. Since the SM asymmetries $A^c_{\rm FB}$ and $A^c_{\rm CP}$ are so tiny, 
the $D^0 - \bar D^0$ oscillation `background' is irrelevant.  

\section{On LHT Scenarios}
\label{LHT}

\subsection{The Flavour in LHT}
\label{LHTPHIL}

The SM predictions presented above leave a large range in rates for these rare transitions, 
where NP could a priori make its presence felt. So-called Little Higgs models mentioned in the Introduction have been studied extensively over the past decade as a possible NP scenario  \cite{LH,SimHiggs}. 
There the Higgs 
boson appears as a pseudo-Nambu-Goldstone boson of a 
spontaneously broken global symmetry. Rather than attempting to solve the hierarchy problem, they 
`delay the day of reckoning' and address a maybe secondary, yet very relevant problem, namely to reconcile the fact that the measured 
values of the electroweak parameters show no impact from NP even on the level of quantum corrections 
with the expectation that NP quanta exist with masses around the 1TeV scale so that they could be produced 
at the LHC. In some of these models, to achieve this program consistently, one needs an additional discrete symmetry \cite{CL1,CL2,Low}, called T parity.   One way of consistently
implementing T parity also requires the introduction of the so-called `mirror' fermions  -- one for each quark 
(and 
lepton) species -- that are {\em odd} under T parity and familywise mass degenerate. This introduces two $3\times 3$ mixing 
matrices $V_{Hd}$ and $V_{Hu}$
neither of which need to be close of the CKM matrix, but they are related to each 
other \cite{VHDVHUCKM}: 
\beq
V_{Hd}^{\dagger}V^{}_{Hu} = V_{CKM}
\label{VVV}
\eeq
Since the CKM matrix does not differ 
too much from the identity matrix, one realizes that LHT contributions exhibit 
a clear correlation of the phases in the charm and strange sector.

In this note we will analyze a subclass of Little Higgs models, namely 
Littlest Higgs Models with T parity (LHT) \cite{L2H,LHTRev}.  In our view they possess several significant strong points: 
\begin{itemize}
\item 
They contain several states with masses that can be below 1 TeV; i.e. those states should be 
produced and observed at the LHC. 
\item 
Compared to SUSY models they introduce many fewer new entities and observable parameters. 
\item 
Their motivation as sketched above lies outside of flavour dynamics. Thus they have not been `cooked up' 
to induce striking effects in the decays of hadrons with strangeness, charm or beauty. 
\item 
Nevertheless they are {\em not} of the minimal flavour violating variety! 
\item 
The impact of LHT dynamics on $K$, $B$ and 
also $D$ transitions has been explored in considerable detail, and potentially sizable effects have  
been identified  \cite{angph,LHTBK,rareBuras, BlankeUpdate,PBR1}.
\item 
Especially relevant for our study is the fact that they can have an observable impact on $D^0 - \bar D^0$ oscillations 
 \cite{DKdual,DMB}. Also sizable indirect 
CP violation can arise in $D^0$ decays \cite{DKdual} very close to the present experimental upper bounds. Having seen such large effects in $\Delta C=2$ transitions coupled with the possibility of the existence of large CP violating phases, one would naturally ask whether it is possible to see the same in $\Delta C=1$ transitions such as $D\to X_u l^+l^-$.
\end{itemize}

\boldmath
\subsection{LHT contributions to $D\to X_u l^+ l^-$}
\unboldmath
As T parity forbids tree level coupling of SM particles with the new T-odd particles, LHT makes it presence felt only through loop contributions from internal mirror fermions and heavy gauge bosons. Unlike in the case of $B$ and $K$, the new T-even heavy top quark does not contribute and hence any new contribution from LHT comes from the T-odd particles. The following are the modifications of the SM functions in Eq.\ref{eq:SM}. The auxiliary functions are defined explicitly in Appendix \hyperref[APPB]{B}. 
\begin{eqnarray}
\nonumber C_1(x)&=&\frac{1}{64}\frac{v^2}{f^2}\left(x S(x)-8x R_2(x)+\frac{3}{2}x+2x F_2(x)\right)\\
\nonumber D_1(x)&=&\frac{1}{4}\frac{v^2}{f^2}\left(D_0(x)+Q_u\frac{1}{2} E_0(x)+Q_u\frac{1}{10} E_0(x^\prime)\right)\\
\nonumber E_1(x)&=&\frac{1}{4}\frac{v^2}{f^2}\left(E_0(x)+\frac{1}{2} E_0(x)+\frac{1}{10} E_0(x^\prime)\right)\\
\nonumber D^\prime_1(x)&=&\frac{1}{4}\frac{v^2}{f^2}\left(D^\prime_0(x)+Q_u\frac{1}{2} E^\prime_0(x)+Q_u\frac{1}{10} E^\prime_0(x^\prime)\right)\\
\nonumber E^\prime_1(x)&=&\frac{1}{4}\frac{v^2}{f^2}\left(E^\prime_0(x)+\frac{1}{2} E^\prime_0(x)+\frac{1}{10} E^\prime_0(x^\prime)\right)\\
\nonumber Y_1(x,y)&=&\frac{1}{64}\frac{v^2}{f^2}\left[S(x)+F_W(x,y)-4(G_Z(x,y)+G_A(x^\prime,y^\prime)+G_\eta(x,y))\right]\\
\end{eqnarray}
where
\begin{eqnarray}
&&x=\frac{m_{Hi}^2}{m_{W_H}^2}=\frac{m_{Hi}^2}{m_{Z_H}^2}\;,\; x'=ax\;,\;a=\frac{5}{\tan^2\theta_W}\nonumber\\
&&y=\frac{m_{l_H}^2}{m_{W_H}^2}=\frac{m_{l_H}^2}{m_{Z_H}^2}\;,\;y'=ay\;,\;\eta=\frac{1}{a}\nonumber\\
\label{eq:def}
\end{eqnarray}
$Q_u=2/3$ is the charge of the up-type quarks, $m_{H_i}$ is the mass of the mirror quark in the $i^{th}$ family and $m_{l_H}$ is the mass of the heavy internal neutrino. The functions $F_W(x,y)$, $G_Z(x,y)$, $G_A(x,y)$ and $G_\eta(x,y)$ are contributions from $WW$, $ZZ$, $AA$ and $ZA$ box diagrams with heavy internal neutrinos respectively. A complete list of Feynman diagrams can be found in \cite{rareBuras}. Since the operator structure is the same in LHT as in SM, the expression for the decay rate and the asymmetries remain the same with the necessary modifications of the Wilson coefficients. QCD corrections to the LHT contributions have been ignored; after all we do not know the model parameters, and these numerical exercises serve to show whether such models can be significant for such 
observables. 

\section{Numerical Findings on LHT Contributions}
\label{RESULTS}
The structure of the mirror fermion sector leaves us with a lot of liberty to choose the parameter space we wish to scan. However, constraints from $B$ and $K$ physics set very stringent limits on the viable parameter space for probing $D$ physics. In what follows, we define an operational parameter space and what effects LHT can bring about in $D\to X_u l^+l^-$.
\subsection{LHT Parameter Space}
\label{LHTPAR}
\begin{figure}[h!]
\includegraphics[width=16cm]{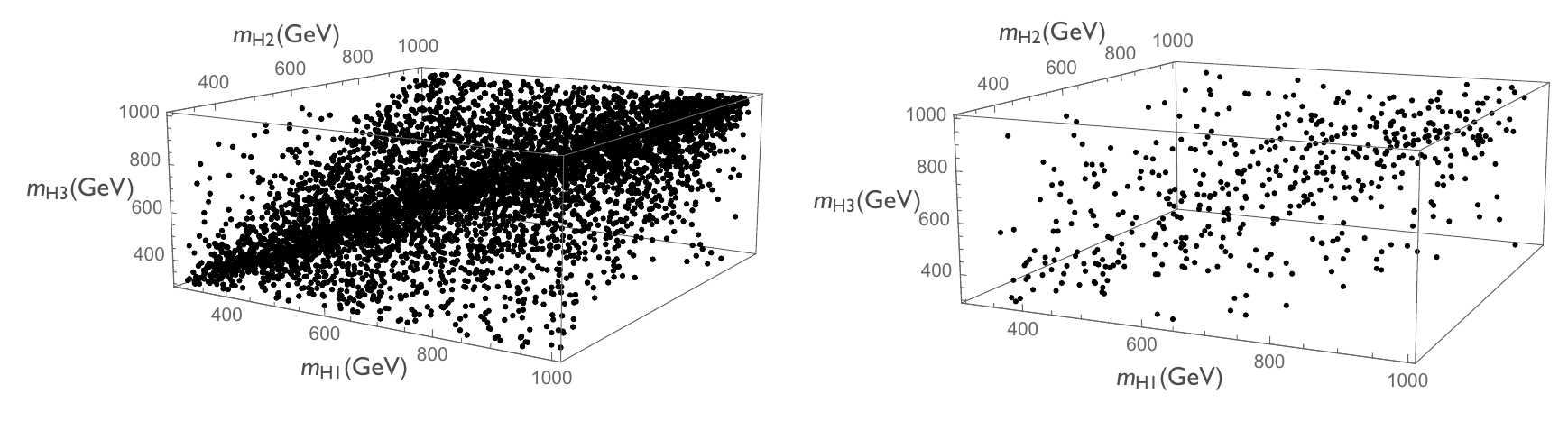}
\caption{Parameter space of the mass of the mirror quarks}
\label{fig:mass}
\end{figure}
\begin{figure}[h!]
\subfigure{
\includegraphics[width=16cm]{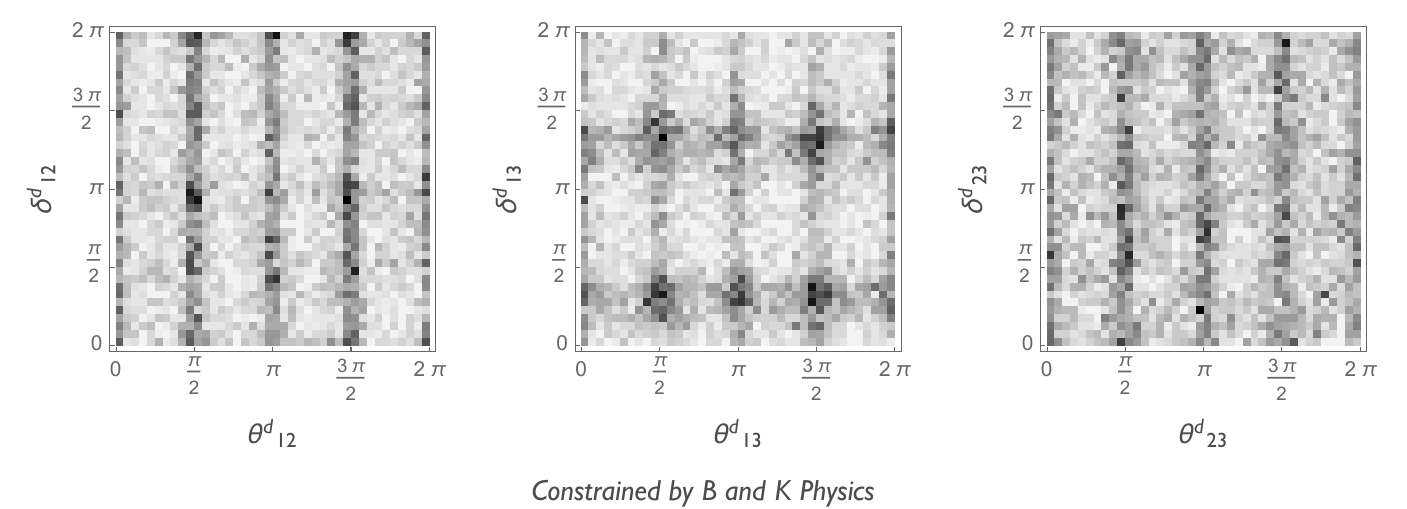}
}
\subfigure{
\includegraphics[width=16cm]{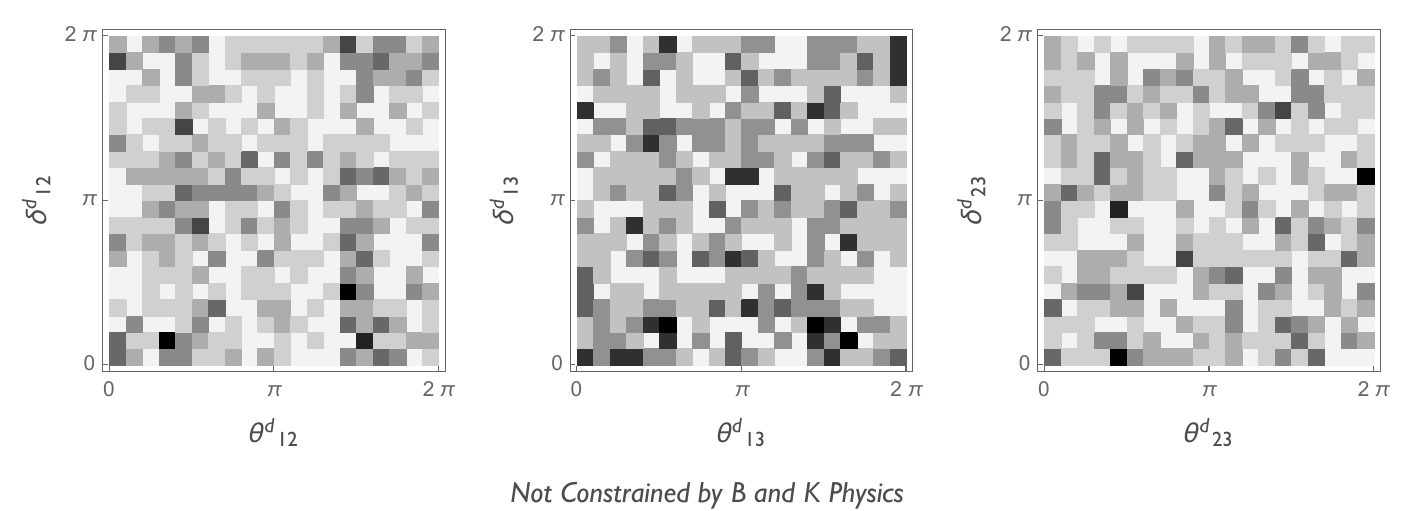}
}
\caption{Histogram of  the parameter space of the angles and phases in $V_{Hd}$. Counts in any bin are represented in grayscale, darker representing higher density }
\label{fig:angles}
\end{figure}
We continue to use the same parameter sets that we used to study the processes $D^0\to \gamma\gamma$ and $D^0\to \mu^+\mu^-$ in our previous work \cite{PBR1}. We vary the mirror fermion masses and the mixing angles and phases over the parameter sets keeping the breaking scale of the non-linear sigma model fixed at 1 TeV\footnote{An analysis of the parameter space can be found in \cite{FPR}.}. The LHT has 20 new parameters of which the ones which will be relevant to us are as follows:
\begin{itemize}
\item  The LHT breaking scale $f=1$ TeV  is fixed by choice.  
\item The masses of the three familywise degenerate T-odd mirror quarks, $m_{H1}, m_{H2}, m_{H3}$ range from 300 to 1000 GeV. 
\item There are three independent mixing angles in 
$V_{Hu}$, $\theta_{12}^{u},\theta_{13}^{u},\theta_{23}^{u}$. 
\item There are three irreducible phases in $V_{Hu}$, $\delta_{12}^{u},\delta_{13}^{u},\delta_{23}^{u}$. 
\end{itemize}
The parameter space used for these analyses is a set that satisfies all experimental constrains from $B$ and $K$ physics. A small parameter set was also used which did not follow such constraints to check whether constraints from $B$ and $K$ physics affects LHT contributions to $D$ physics. However, even the parameter set that is not constrained does not have large mass hierarchies in the mirror fermion sector.

The mass spectrum for both the parameter sets is illustrated in Figs.\ref{fig:mass}. Using Eq(\ref{VVV}), the angles and phases of $V_{Hu}$ were calculated from those of $V_{Hd}$ and hence were constrained by $B$ and 
$K$ physics too for the first parameter set and not so for the second. Histograms of the parameter space of the angles and phases are shown in Figs.\ref{fig:angles}. The angles and phases are family-wise paired.

\boldmath
\subsection{Impact on $\Gamma_{\rm SD}\left(D\to X_u l^+ l^-\right)$}
\unboldmath
\label{RESBR}
\begin{figure}[h!]
\centering
\subfigure[Constrained by $B$ and $K$ Physics]{\includegraphics[width=16cm]{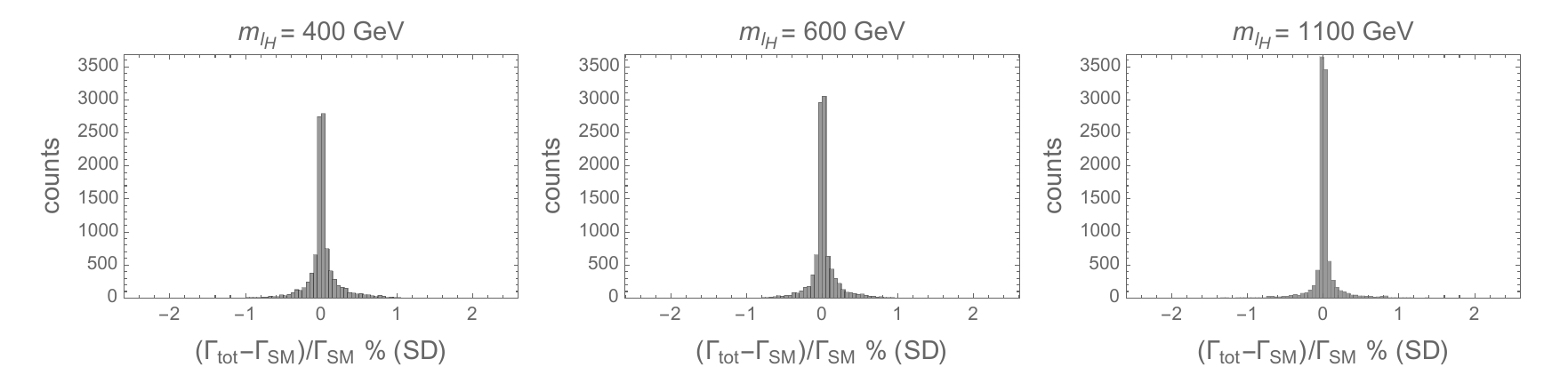}
\label{fig:BRCC}}
\subfigure[Not Constrained by $B$ and $K$ Physics]{\includegraphics[width=16cm]{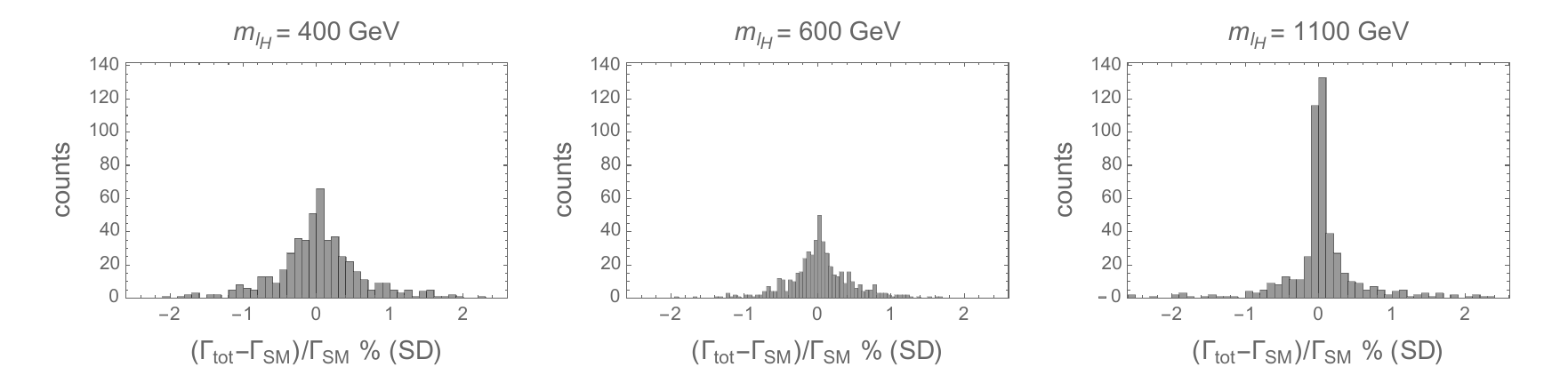}
\label{fig:BRNC}}
\caption{Percentage change of the SM SD contribution to $\Gamma_{\rm SD}(D\to X_ul^+l^-)$ due to LHT effects.}
\label{fig:BRLHT}
\end{figure}
As we have seen in Sec.\ref{DECAY}, the dominant SM contributions to $\Gamma_{\rm SD}\left(D\to X_u l^+ l^-\right)$ are through the $\gamma$ penguin in the $O_9$ operator and some from the mixing of $O_{1-6}$ with $O_9$.  The subdominant contributions come from the two loop $O(\alpha_s)$ term in $C_7$. Hence part of the dominant effect and the subdominant effect both come from QCD corrections. All other contributions are smaller by orders of magnitude. We saw in our previous work on  $D\to \gamma \gamma$ and $D\to \mu^+\mu^-$ \cite{PBR1} that LHT is capable of producing large enhancements through box diagrams with internal heavy fermions and heavy gauge bosons just as in $D^0-\bar{D}^0$ oscillation \cite{DKdual} and somewhat moderate enhancements to $Z_L$ penguins. However, the enhancement to effective $\gamma$ vertices are tiny compared to the SM contributions. 

In $D\to X_ul^+l^-$ we confirm our previous conclusions. The decay rate was calculated for three different internal heavy neutrino mass of 400 GeV, 600 GeV and 1100 GeV. We see very tiny change in both the differential decay rate and the integrated decay rate at $O(1\%)$ as can be seen from Fig.\ref{fig:BRLHT} in which the abscissa represents the percentage enhancement to the SD decay rate after inclusion of LHT. Removing constraints from $B$ and $K$ physics does not make much difference either as can be understood from comparing Fig.\ref{fig:BRCC} and Fig.\ref{fig:BRNC}. Our result is different from what was found in \cite{FajferLH} which used the Littlest Higgs model with{\em out} T parity. Without T parity the  $SU(2)$ custodial symmetry protecting the electroweak $\rho$ parameter is explicitly broken at scales below 4 TeV  \cite{Han} which is brought about by the 
$U(1)_H$ gauge boson. This model allows large tree level FCNC mediated by the coupling of the $Z_H$ and $A_H$ heavy gauge bosons with the SM quarks. The enhancements seen in \cite{FajferLH} is because of the existence of these tree level FCNCs which are absent from LHT. Hence, LHT makes almost no changes to $C_7$ or $C_9$ above SM contributions and hence fails to enhance the decay rate. This essentially means that any LHT contribution to the decay rate fails to significantly dent SM SD contributions and is completely swamped by SM LD effects.

\boldmath
\subsection{Impact on $A^c_{\rm FB}$ and $A^c_{\rm CP}$ and $A^{\rm CP}_{\rm FB}$}
\unboldmath
\label{RESAFBACP}
\begin{figure}[h!]
\centering
\subfigure[Constrained by $B$ and $K$ Physics]{\includegraphics[width=16cm]{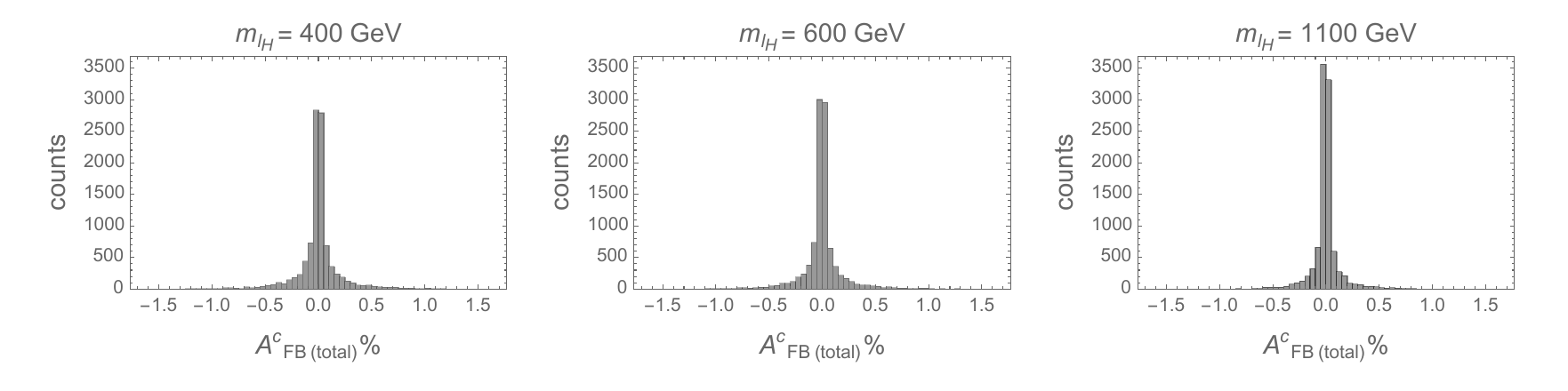}
\label{fig:AFBCC}}
\subfigure[Not Constrained by $B$ and $K$ Physics]{\includegraphics[width=16cm]{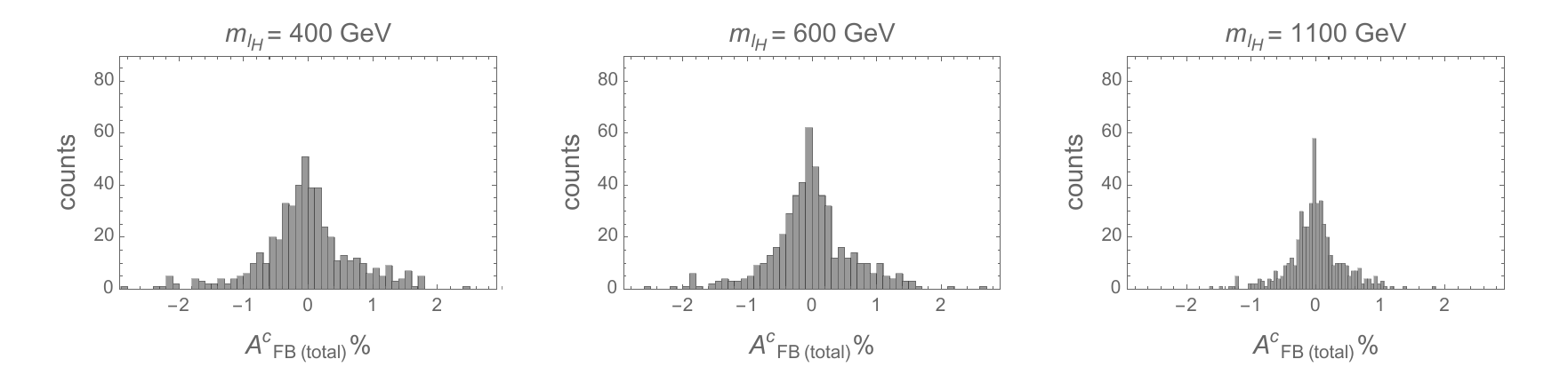}
\label{fig:AFBNC}}
\caption{$A^c_{\rm FB}$ after including LHT effects.}
\label{fig:AFBLHT}
\end{figure}
\begin{figure}[h!]
\centering
\subfigure[Constrained by $B$ and $K$ Physics]{\includegraphics[width=16cm]{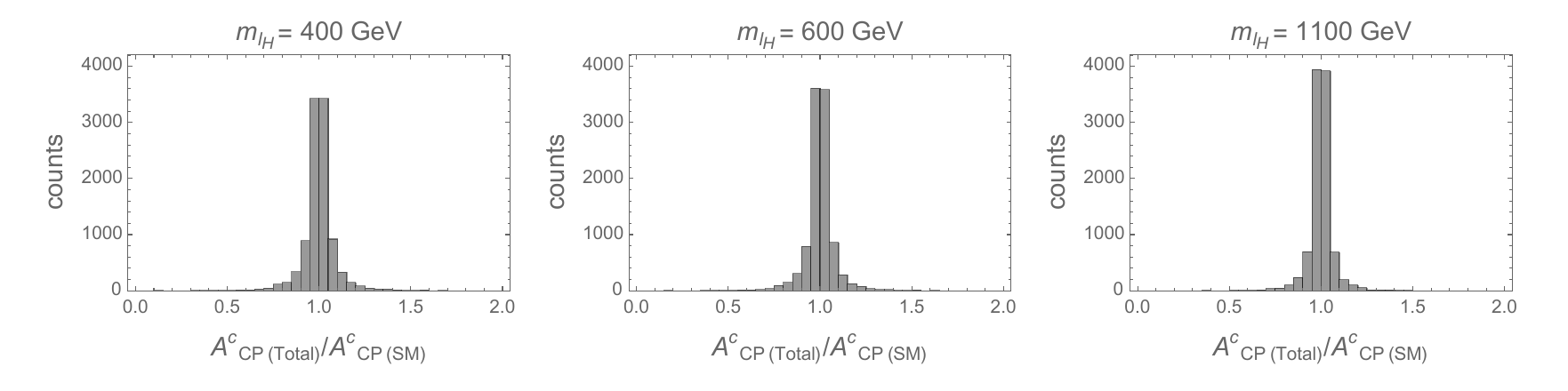}
\label{fig:ACPRelCC}}
\subfigure[Not Constrained by $B$ and $K$ Physics]{\includegraphics[width=16cm]{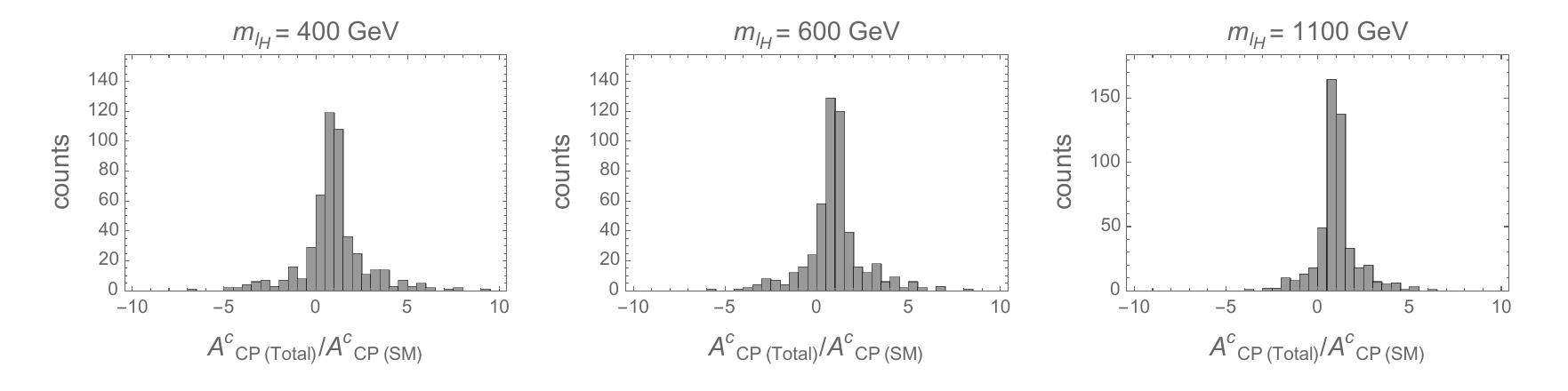}
\label{fig:ACPRelNC}}
\caption{Enhancement to $A^c_{\rm CP}$ over SM after including LHT effects.}
\label{fig:ACPRelLHT}
\end{figure}
\begin{figure}[h!]
\centering
\subfigure[Constrained by $B$ and $K$ Physics]{\includegraphics[width=16cm]{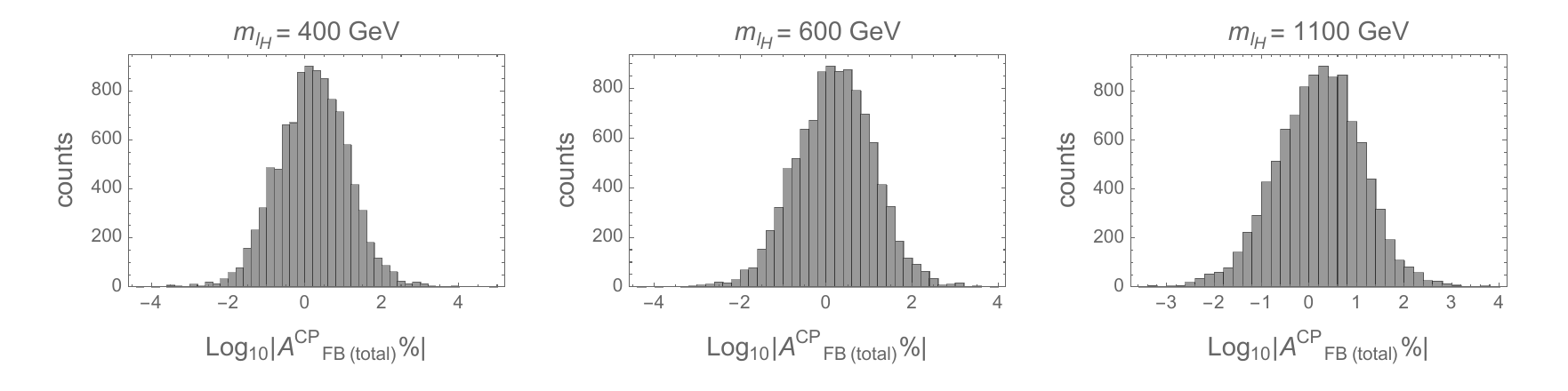}
\label{fig:AFBCPCC}}
\subfigure[Not Constrained by $B$ and $K$ Physics]{\includegraphics[width=16cm]{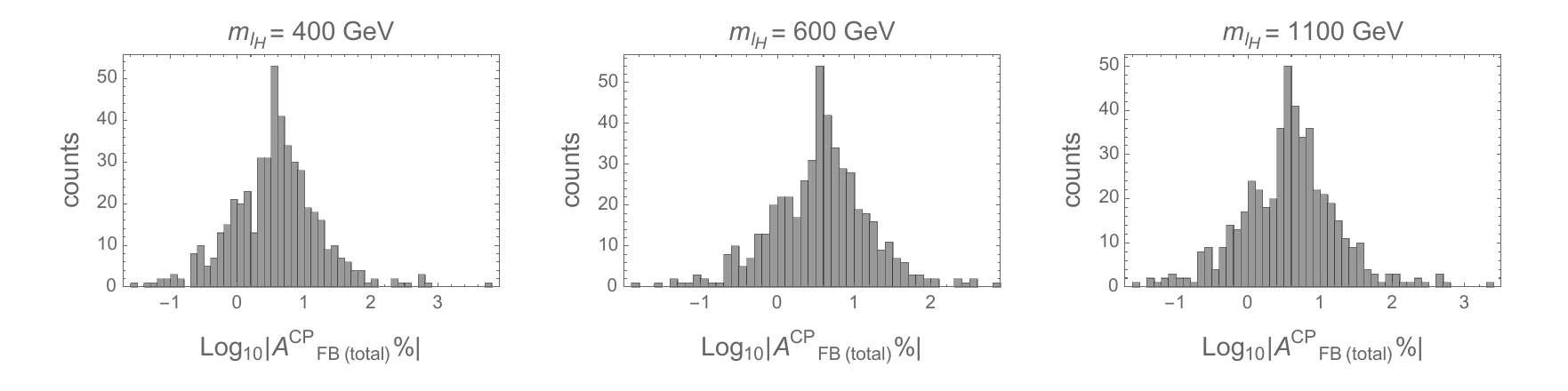}
\label{fig:AFBCPNC}}
\caption{Large enhancements in $A^{\rm CP}_{\rm FB}$ after including LHT effects.}
\label{fig:AFBCP}
\end{figure}

As we have seen earlier in Sec.\ref{ASYMM}, the SM contribution to  $A^c_{\rm FB}$ is all but nonexistent due to tiny SM contributions to $C_{10}$. In LHT $C_{10}$ gets enhanced by orders of magnitude which brings about a large enhancement in $A^c_{\rm FB}$. This effect is similar to what was observed in \cite{FajferLH} but comes from box diagrams involving T-odd heavy internal degrees of freedom rather than tree level FCNC. It is also commensurate with the enhancement we found in SD contribution to 
$\Gamma(D^0\to \mu^+\mu^-)$ \cite{PBR1} from LHT. This is due to the fact that SD contribution to 
$\Gamma(D^0\to \mu^+\mu^-)$ comes from $O_{10}$ and $A^c_{\rm FB}$ is highly sensitive to the same. However, even with such a large enhancement, the absolute value $A^c_{\rm FB}$ after including the LHT enhancement can at most be of $O(0.5\%)$ as can be seen from 
Fig.\ref{fig:AFBLHT}. Studying Fig.\ref{fig:AFBCC} and Fig.\ref{fig:AFBNC} shows that removing constraints from $B$ and $K$ physics creates large {\em relative} enhancements -- it can be as much as $1 \%$ (or rarely more) -- but they do not enhance it to sizable absolute effects.

On the other hand $A^c_{\rm CP}$ depends mostly on $C_7$ and $C_9$, which we have already seen suffers almost no enhancement from LHT. However, as $A^c_{\rm CP}$ is quite sensitive to the phases in these coefficients, it gets enhanced by a few factors over the SM value and can be as large as four times the SM value. This can be seen from Fig.\ref{fig:ACPRelLHT} where we plot the ratio of the total $A^c_{\rm CP}$ including the LHT enhancements to the SM value of the same. However, this still keeps $A^c_{\rm CP}$ at $O(10^{-4}) - O(10^{-3})$ and hence the absolute measure of the CP asymmetry is still experimentally challenging. The unconstrained parameter set allows for  slightly larger enhancements to $A^c_{\rm CP}$ but is limited to almost  the same order of magnitude.

The contributions from LHT models can enhance $A^{\rm CP}_{\rm FB}$ so much as to bring it up to possibly measurable values. As pointed out in Sect.\ref{ASYMM}, $A^{\rm CP}_{\rm FB}$ is sensitive to any phase in $C_{10}$. LHT 
can not only enhance the magnitude of $C_{10}$ by orders of magnitude, but also brings about the possibility of existence of a very large phase in it. For $C_7$ and $C_9$ the effect is dominated by phases from QCD radiative corrections. 
The existence of this large phase and the tangential dependence of $A^{\rm CP}_{\rm FB}$ on it results in the huge enhancement that we see in $A^{\rm CP}_{\rm FB}$ as illustrated in Fig.\ref{fig:AFBCP} where we plot the total $A^{\rm CP}_{\rm FB}$ after the inclusion of LHT effects. This is commensurate with what was observed in \cite{DKdual} for CP violation in $D^0-\bar{D}^0$ oscillations. For both the constrained 
(Fig.\ref{fig:AFBCPCC}) and unconstrained (Fig.\ref{fig:AFBCPNC}) sets more than $10\%$ of the parameter set can produce asymmetries of $O(10\%)$ or greater!  As explained 
in Sect.\ref{ASYMM}  we relate $A_{\rm FB}^{\rm CP}$ with SD contribution to $D \to l^+l^-X$, which amounts to a branching ratio of $1.5 \times 10^{-9}$. With a sample of $10^{13}$ $D$ mesons 
NP intervention should be measurable for $A_{\rm FB}^{\rm CP} > 5 \%$. Such effects could be within the reach of to the currently  running LHCb experiment and ones like the planned and approved SuperB Collaboration.

\boldmath
\subsection{Correlation between $A^c_{\rm FB}$ and $\Gamma_{\rm SD}\left(D\to X_u l^+ l^-\right)$}
\unboldmath
\label{FCNC}
\begin{figure}[h!]
\includegraphics[width=17cm]{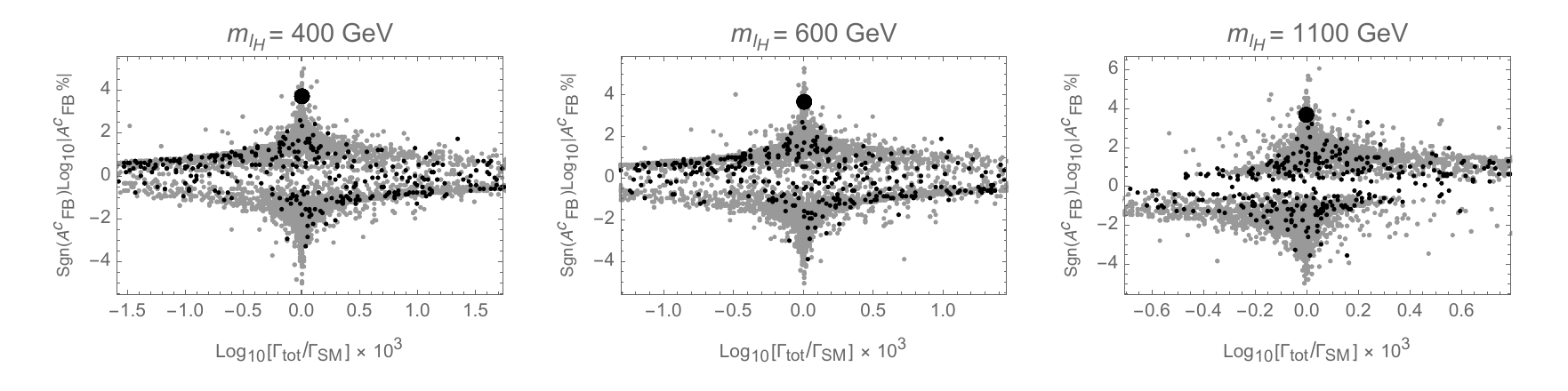}
\caption{Correlation between $A^c_{\rm FB}$ and $\Gamma_{\rm SD}\left(D\to X_u l^+ l^-\right)$. The gray points represent the constrained set and the black ones for the unconstrained one. The big black spot represents the SM values.}
\label{fig:Corr}
\end{figure}

In Fig.\ref{fig:Corr} we plot the correlation between $A^c_{\rm FB}$ and $\Gamma_{\rm SD}\left(D\to X_u l^+ l^-\right)$. 
These plots look at first chaotic, yet a careful (and time consuming) study reveals a pattern. For `low' heavy neutrino masses   
$m_{l_H}= 400$ GeV, $600$ GeV the LHT parameter sets that enhance $\Gamma_{\rm SD}\left(D\to X_u l^+ l^-\right)$ 
can produce positive $A^c_{\rm FB}$ in some regions and negative ones in others; on the other hand sets 
decreasing $\Gamma_{\rm SD}\left(D\to X_u l^+ l^-\right)$ can 
also produce positive and negative $A^c_{\rm FB}$, but in others regions. Those `low' heavy neutrino masses are within the range of masses used for the mirror quarks in this study. However, for  $m_{l_H}=1100$ GeV those LHT parameters that increase the SD 
branching ratio produce mostly a negative $A^c_{\rm FB}$, while sets decreasing the SD 
branching ratio lead mostly a positive $A^c_{\rm FB}$.  At this mass the mirror neutrino is  heavier than any of the mirror quarks. 

This behavior can be understood quite well. Dependence on the heavy neutrino mass exists only in the box diagrams. The sign of $A^c_{\rm FB}$ depends on the sign of $C_{10}$ which depends on the relative size of the mirror quarks and heavy neutrino masses which comes from box diagrams. At less than 1 TeV, this can go either ways with the neutrino being either heavier or lighter than one or more of the mirror quarks with slightly greater chances of being lighter than them. At above 1 TeV the heavy neutrino is always heavier than the mirror quarks used. Also enhancements to 
$\Gamma_{\rm SD}\left(D\to X_u l^+ l^-\right)$ are quite sensitive to the box diagrams through both $C_{10}$ and $C_{9}$. This leads to the sharp change in the correlation we see in Fig.\ref{fig:Corr}. 

We do not see any such correlation in $A^c_{\rm CP}$ vs. $\Gamma_{\rm SD}\left(D\to X_u l^+ l^-\right)$ or $A^c_{\rm FB}$ vs. $A^c_{\rm CP}$; after all $A^c_{\rm CP}$ is blind to $C_{10}$ and hence  less sensitive to the box diagrams as in this case they manifest themselves only through $C_{9}$. Furthermore we see no correlation in 
$A^{\rm CP}_{\rm FB}$ vs. $\Gamma_{\rm SD}\left(D\to X_u l^+ l^-\right)$ or $A^{\rm CP}_{\rm FB}$ vs. $A^c_{\rm FB}$ 
since $A^{\rm CP}_{\rm FB}$ is not affected by the magnitude of $C_{10}$, but rather by the CP violating phase in it; that  information is lost in both $\Gamma_{\rm SD}\left(D\to X_u l^+ l^-\right)$ and $A^c_{\rm FB}$. Lastly, we see no correlation between $A^c_{\rm CP}$ and $A^{\rm CP}_{\rm FB}$ even though both are CP violating parameters. This reinforces our earlier statement that the sources of CP violation are distinct in these parameters with the former coming from phases in $C_7$ and $C_9$ and the latter coming from a phase in $C_{10}$.

\section{Further Insights into FCNCs in LHT-like Models}
\label{FCNC}
Our work on the impact of LHT on $D^0\to \gamma \gamma/ \mu^+\mu^-$ \cite{PBR1} had lead us to some general conclusions on the structure of FCNCs within a LHT-like framework. As defined previously, this framework contains
\begin{itemize}
\item A second sector of fermions that are an exact copy of the SM ones.
\item  Mass mixing matrices which are unitary and loosely connected to $V_{CKM}$ (Eq.\ref{VVV}).
\item Possible large angles and phases in the mass mixing matrices.
\item Possible large hierarchies in the masses of the mirror quarks.
\item A symmetry, like T parity, segregating the NP sector from the SM sector, hence forbidding tree level FCNC.
 \end{itemize}
We shall, after further investigation, relax the second condition to:
\begin{itemize}
\item Mass mixing matrices that are constrained by a relationship between the one(s) connecting the new Up-type quarks with the SM down-type quarks to the one(s) connecting the new Down-type quarks with the SM up-type quarks.
\end{itemize}

 In general, FCNCs are a very sensitive probe to the details of the flavour structure of both the SM and any NP models as they highlight not only mass hierarchies within a theory but also are sensitive to phases within the same. Moreover, it is possible to disentangle the effect of phases and fermionic mass hierarchies on FCNCs in a model independent way if we have access to more observables. 
 
 $D\to X_u l^+l^-$ wins over $D^0\to \gamma \gamma$ and $D^0\to \mu^+\mu^-$ by leaps and bounds in this respect. While $D^0\to \gamma \gamma$ is sensitive mostly only to $O_7$ and $D^0\to \mu^+\mu^-$ is sensitive only to $O_{10}$,  $D\to X_u l^+l^-$ is not only sensitive to all of that but also to many more. Moreover, the final state being a three body final state, this channel can also be probed through forward-backward and CP asymmetries hence opening the possibility of probing phases in any model, too. 

 As we noted above, the SD contribution to branching fractions is dominated by the photonic penguin in $O_9$ while $A^c_{\rm FB}$ is highly sensitive to $O_{10}$, $A^c_{\rm CP}$ to the mixing between $O_7$ and $O_9$ and 
$A^{\rm CP}_{\rm FB}$ to the phase in $C_{10}$. Studying 
 $D\to X_ul^+l^-$ in a sufficiently precise way, we can learn the impact of several operators and then comment 
 on other rare decays. Logically we should have started our analysis with $D\to X_ul^+l^-$ and then applied 
 our findings to the simpler cases of two-body rare decays. Instead we started with our analysis of two-body rare decays, from which we extracted some conjectures; they happened to be correct in more general theoretical situations.
 
 From our results we see that, through LHT dynamics, $A^c_{\rm FB}$ gets orders of magnitude enhancement through the enhancement of 
 $C_{10}$ which is commensurate with the orders of magnitude enhancement that we had seen in the SD contribution to $D^0\to \mu^+\mu^-$ which was ultimately overshadowed by the LD contribution to the branching fraction. We see almost no enhancement to $C_7$ which agrees with the lack of enhancement that we noted in $D^0\to \gamma\gamma$. It is not that $C_7$ gets absolutely no enhancement from LHT. The purely electroweak part of $C_7$ does get moderately enhanced. However, this enhancement is completely overshadowed by the SM two-loop $O(\alpha_s)$ QCD correction, something which is peculiar to the $D$ meson system and not seen in the $B$ system. In addition, we also see almost no enhancement to $C_9$ coming from the fact that photonic penguins are not enhanced by LHT, which results in a lack of enhancement to the branching fraction of $D\to X_u l^+ l^-$.
 
 Intuitively one might think that the situation for $A^c_{\rm CP}$ should be different as it is sensitive to phases in $C_9$ and $C_7$, and LHT allows for large phases in the mixing matrices. In a purely electroweak SM scenario this would have been true. However, $A^c_{\rm CP}$ gets an unusual boost within the SM from the unusual two-loop 
 $O(\alpha_s)$ QCD correction, something that LHT can barely overcome. Coupled to the fact that $C_9$ does not gain much from LHT, enhancements to $A^c_{\rm CP}$ fail to impress. The validity of the previous statement is further tested when we see orders of magnitude enhancement in $A^{\rm CP}_{\rm FB}$ as it gets a boost from the introduction of a large phase from LHT while it is completely clean of phases from the SM.
 
So what does this tell us about LHT-like models and their effects in flavour physics? In principle in any such model, large angles and phases in mass mixing matrices and large fermionic masses and sharp hierarchies amongst them are possible. However, one has to satisfy the experimental constraints that we already have in $B$ and $K$ physics. At this point one will have to choose between large angles and phases or huge hierarchies in the fermionic masses as experimental data already tell us that the extra fermions have to be heavy. Making such a choice automatically limits the size of NP intervention in yet unobserved FCNC processes, specially in $\Delta F=1$ processes even if 
$\Delta F=2$ processes can escape these limits and absorb NP contributions. However, an exception to this rule occurs when these large phases from New Physics are laid bare and have purely SM electroweak effects to compete with as in the case of $A^{\rm CP}_{\rm FB}$ where we see large effects even in a parameter which is a measure of a 
$\Delta F=1$ process. Hence, let us have a more thorough look at the diagrammatic details of NP intervention from LHT-like models.
  
 \section[On Boxes and Penguins in New Physics]{On Boxes and Penguins in New Physics\footnote{and Seagulls too, but for now we shall ignore them.}}
 \label{BOXPEN}
  Due to CPT symmetry, CP violation can enter only through complex effective couplings. In the SM they can 
 arise only for the weak boson couplings to quarks as described by CKM matrices. They are 
 necessarily unitary, since all quark masses are given by a single VEV of a neutral Higgs multiplied by 
 numbers, not a matrix. The concept of `weak universality' was first put forward by Cabibbo in 1967 \cite{NICOLA}. 
Afterwards it was scrutinized experimentally. Later it was understood if the weak forces are embedded -- as it applied to the SM -- in a single non-abelian gauge theory, weak universality has to hold. There is the Singular Value Decomposition theorem which tell us that the matrices relating mass and flavour 
left $Up$ and $Down$ quarks are unitary and therefore their matrix product -- the CKM matrix -- is also unitary. 
For $N=2$ families their phases can be transformed away, for $N=3$ there is one irreducible phase that is 
therefore observable. CP violation can surface in processes where quarks from three families can contribute 
as real or virtual entities. The latter happens in SM, since FCNCs arise effectively through quantum 
corrections, namely box and penguin diagrams. There are box diagrams in only one kind, the $WW$ box. Penguins come in three varieties: 
the $Z_L$, the $\gamma$ and the gluon penguins are particularly essential for direct CP violation. 

For $N$ families the unitarity of $V_{CKM}$ translates into\footnote{This triangle relation holds when $\{j,k\}$ belong to the up-type sector. If they belong to the down-type sector we have $\lambda_i=V^*_{ij}V^{}_{ik}$ instead with $\{i\}$ coming from the up-type sector. All other arguments hold true.} 
\beq
 \sum_{i=1}^N \lambda_i=\delta_{jk},\text{ with } \lambda_i=V^*_{ji}V^{}_{ki}
 \label{eq:uni}
\eeq
with $|\lambda _i| > 0$ for $i=1,2... N$ families with $\{i\}$ coming from the down-type sector and $\{j,k\}$ coming from the up-type sector -- i.e., a triangle equation in the complex plane: CP violation 
arises at the mass generation. If two of the quark were 
mass degenerate, CP invariance would survive mass generation -- yet it is not the case on our world. 

If the SM contained a fourth family, it could -- and probably would -- have two more observable phases 
with more {\em independent} CP asymmetries.  

The very fact that in the SM one has to study transitions where quarks from three families contribute reduces the 
number of `interesting' cases for CP violation very significantly. The observation that absolute values of off-diagonal elements of $V_{CKM}$ are small, 
even tiny, in most such cases are experimentally `challenging' at least -- except in beauty decays, since leading 
decays are so suppressed and therefore make them experimentally challenging for a different reason. 
It was expected that absolute values of the off-diagonal elements are small, but it was surprising that $|V_{bc}|$, 
$|V_{bu}|$ etc. are so tiny, namely much smaller than $|V_{su}|$.  Originally it was conjectured -- based 
on no good theoretical reason -- that also the CKM phase is small. Now we know that the CKM phase is 
not small. Therefore NP is likely to exhibit also sizable phases -- like LHT. 

The size of CP asymmetries depends on the phases, the absolute values of the quark mixing matrix elements and the masses 
of the internal virtual gauge bosons and fermions and their mass hierarchies. Let us analyze to what degree each criterion  can be fulfilled in this class of NP models we call LHT-like for box and Penguins. 
The range of masses of mirror quarks and $W_H$ and $Z_H$ is rather limited -- for two very different reasons: 
(a) None have been found up to $\sim 200$ GeV. (b) They could be much higher like 10 TeV or even 100 TeV. 
Then they could not be directly produced at the LHC. Therefore we stop our analysis at the 1-2 TeV scale. 
In that case some of the new states could mix significantly. The gauge boson mass is also of the 
 $O(1\, {\rm TeV})$. Large angles and phases are possible but limited by experimental constraints from $B$ and $K$ physics. Also, quite important to our analysis, hierarchies in fermionic masses are very small, a lot smaller that what we find in the SM.

In LHT-like models the operators with new degrees of freedom are similar to the SM SD ones. Hence they scale similarly to the SM SD operators which has been established for decades now \cite{InamiLim} and has been used extensively to qualitatively judge the size of flavour dynamics ever since. Let us reexamine this scaling behavior. Of course, this scaling is only an approximation of the detailed formfactors, however, most of the time such approximations are enough to estimate the size of many effects in flavour physics.The mass hierarchy in the new fermionic sector can be established with
 \begin{eqnarray}
 m_i^a=m_1^ah_i^a\phantom{xx}\forall i=1 \ldots  N
 \label{eq:hier}
 \end{eqnarray}
 Here $a$ distinguishes amongst the members of each family which defines the the sector to which the fermion belongs and it is {\it not} summed over. It is possible to set $h_1^a\ne 1$ and choose $m_1^a$ to be any finite mass representative of the phyics in consideration. The formfactors in flavour physics can be approximately expressed as

\begin{eqnarray}
F(x)=f_n(x)(\log(x))^m\text{  where,  }f_n(x)=x^n, n\in \mathbb{Z}, m=0,1
\label{eq:form}
\end{eqnarray}
Ignoring QCD (or QCD-like) corrections, any matrix elements involving the processes mentioned above are of the form
\begin{eqnarray}
\mathcal{M}\sim\sum_{i=1}^N \lambda_i F(x_i)
\label{eq:sum}
\end{eqnarray}
Here, $x^a_i=(m_i^{a}/m_G)^2$ is the commonly used square of the ratio of the internal fermion mass to the mass scale of the massive gauge bosons.  The superscript $a$ has been dropped as the matrix element usually involves just one sector. Using the unitarity relation in Eq.\ref{eq:uni}, it can be shown that under the hierarchy defined in Eq.\ref{eq:hier}, Eq.\ref{eq:sum} scales as
\begin{eqnarray}
\sum_{i=1}^N \lambda_i F(x_i)=f_n(x_1)\sum_{i=1}^N\lambda_i F(h_i^2)+F(x_1)\sum_{i=1}^N\lambda_i f_n(h_i^2)
\end{eqnarray}
Note that the mass scale of the fermions does not suffer from unitarity suppression but the  hierarchy does. In other words, the only two ways of getting large matrix elements are either to start at a very high mass scale for fermions or to build a very strong hierarchy that will illude the unitarity suppression, or both.

Now, let us have a look at boxes and penguins. 

\begin{description}
  \item[Box diagrams] \hfill \\
Box diagrams scale as $x$ for both large and small $x$. Setting $n=1$ and $m=0$ in Eq.\ref{eq:form} we see that contributions from these will scale as
\begin{eqnarray}
\mathcal{M}\sim x_1\sum_{i=1}^N\lambda_i h^2_i
\end{eqnarray} 
This clearly tells us that box diagrams are sensitive to both mass scales and large hierarchies. Hence any NP model containing either fermions with masses comparable or greater than the gauge bosons or having a large hierarchy amongst the families or both will make large contributions through box diagrams.
  \item[$Z_L$ Penguins] \hfill \\
  $Z_L$ Penguins scale as $x \log(x)$ for small $x$ and as $x$ for large $x$. In the regime in which they scale as $x$ the conclusion is the same as above. Setting $n=1$ and $m=1$ in Eq.\ref{eq:form}, we see that if, for some reason, NP offers small $x$, the contribution will scale as
  \begin{eqnarray}
\mathcal{M}\sim x_1\sum_{i=1}^N\lambda_i h_i^2\log\left(h^2_i\right)+x_1\log\left(x_1\right)\sum_{i=1}^N\lambda_i h_i^2
\end{eqnarray}
Hence, in addition to the conclusions drawn for large $x$, $Z_L$ Penguins will get large contributions from NP if the fermionic masses are light compared to the gauge boson mass and shows large hierarchies. However, such scenarios will be very unusual if not unheard of in the NP models currently under consideration.
  \item[$\gamma$ and chromomagnetic Penguins] \hfill \\
  Electromagnetic and chromomagnetic penguins scale as $x$ for small values of $x$ but are asymptotic to a constant for large values of $x$. Hence they can only benefit from large hierarchies at fermionic mass scale smaller than the gauge boson. Even if NP has large fermionic masses, even with large hierarchies to offer, it will show up only as moderate enhancements in this class of Penguins. \\
 
  On the other hand, photonic Penguins scale as $\log(x)$. Hence, $n=0$ and $m=1$ in Eq.\ref{eq:form}, we see its contribution to matrix elements will scale as
  \begin{eqnarray}
\mathcal{M}\sim \sum_{i=1}^N\lambda_i \log\left(h^2_i\right)
\end{eqnarray}
NP intervention will fail to produce any enhancement in photonic Penguins unless it has very sharp hierarchies to offer in its fermionic masses. Large mass scales will have no effect on photonic Penguins.
\end{description}

A look at the mass spectrum of the heavy fermions, Fig.\ref{fig:mass}, which we used in the previous study 
\cite{PBR1} and in the present one, show that we have heavy mass scales in the spectrum, but not large hierarchies. As a result, we get large enhancements to purely electroweak processes which involve box diagrams and $Z_L$ penguins. Such is the case for $D^0\to \mu^+\mu^-$  and $A^c_{\rm FB}$ where we see orders of magnitude enhancement to SM SD rates and also in $D^0-\bar{D}^0$ oscillations. $A^{\rm CP}_{\rm FB}$ too benefits from this effect. For electromagnetic penguins and chromomagnetic penguins, NP intervention can only be moderate. This explains the small enhancements that we saw in $D^0\to \gamma \gamma$ which is primarily sensitive to $O_7$. However, in $D\to X_ul^+l^-$, the dominant contribution, by orders of magnitude is the photonic penguins in $O_9$ if we ignore QCD corrections which are anyways blind to NP. Even if LHT manages to enhance the other contributions, it falls short of the SM contribution to $C_9$. Even the LHT contributions to the box diagrams and $Z_L$ Penguins in $O_9$ fails to overcome the SM contribution to the photonic Penguins. Moreover, the photonic Penguins do not see much NP intervention as we have already argued.

 This gives us a qualitative way of understanding why LHT fails to enhance processes driven by $\Delta C=1$ 
 dynamics involving the $D$ meson system unless QCD effects are completely absent. This argument can potentially be extended to any other model which shares the same flavour structure as LHT, what we have previously defined as LHT-like. Of course, we always have to keep in mind that we are heavily constrained from $B$ and $K$ physics and that limits us in the parameter space that we can choose to work with.

\section{Conclusions}
\label{CON}

Charm hadrons stable under strong forces were predicted to keep the SM consistent with the observed suppression 
of strangeness changing neutral currents, as were their preference to decay into strange hadrons with decays into
non-strange hadrons being Cabibbo suppressed and to preserve renormalizability; they were even found in the expected 
mass range. Therefore decays of charm hadrons were hardly seen as worth probing for manifestations of NP.  Furthermore 
one realized that kaon and even more, beauty hadrons could clearly  exhibit NP signals, since their leading 
SM decays are Cabibbo and KM suppressed. A minority of authors argue that the probe for 
NP should not be given up in charm decays, since the SM weak phenomenology is `dull' and one need `only' much more statistics there. 

The case was somewhat strengthened after the observation of $D^0 - \bar D^0$ oscillations, although it is not 
outside some SM estimates. Furthermore it was found that a class of NP models like LHT that is motivated 
from outside the flavour dynamics, can yet produce a sizable contribution to the observed 
$D^0 - \bar D^0$ signal and can create much stronger indirect CP violation \cite{DKdual}.  

$D\to X_u l^+l^-$ is a good laboratory for NP  as it is rich in its operator structure and involves almost anything that FCNCs have to offer, including direct CP violation. However, we find that LHT fails to create a significant dent to SM SD contributions to this channel other than significantly enhancing $A^c_{\rm FB}$ and $A^{\rm CP}_{\rm FB}$ for reasons we have explained above. Moreover, SM LD contribution dominates over both SM SD and LHT contributions to the branching fractions. We reiterate the conclusion that we had come to in our previous work \cite{PBR1}: while LHT can contribute significantly to $\Delta C=2$ processes, it fails to dominate in $\Delta C=1$ processes with orders of magnitude enhancements unless the contribution appears through box diagrams and are bereft of relatively large QCD effects.

We also go ahead and have a second look at what we previously defined as LHT-like models:  again we found 
that certain conclusions can be drawn about NP's flavour structure by studying weak decays independent of the other details of the underlying model. Absence of large hierarchies and unitarity of the new mass mixing matrices within  such NP models heavily  limits  the new FCNCs. Also, we have shown  that experimental limits in $B$ and $K$ physics can be directly ported to give constraints on NP intervention in $D$ physics. 

During our analysis we developed a more general conjecture: if NP models affect the dynamics of both the up-type and down-type quarks in a tightly correlated way -- as shown in Eq.(\ref{VVV}) in the case of LHT -- they will mostly fail to contribute significantly in $\Delta C=1$ dynamics even if they play a major role in $\Delta C=2$ processes. 
For constraints from $\Delta S=1$ and $\Delta B=1$ reactions will suppress $\Delta C=1$ coupling 
greatly, since sensitivity for NP is often greater in $K$ and $B$ decays because of their leading 
SM transitions are Cabibbo or KM suppressed; otherwise it would oversaturate for $\Delta C=2$ effects. The crucial feature on the 
right hand side of Eq.(\ref{VVV}) is only that the CKM matrix is very close to the identity matrix; the same conclusion 
should apply for any matrix in the right hand side that is close to the identity matrix. Of course, the validity of this conjecture is yet to be tested in other $\Delta C=1$ processes like 
direct CP violations in nonleptonic charm decays, keeping in mind that charm transitions could still produce surprises 
for us -- and about SM's `ability' to cope with them. In the event that future experiments reveal a clear manifestation of NP through enhancements in the $\Delta C=1$ processes, it is unlikely that any LHT-like model can be a `culprit'. 

\section{``Postmortem"} 

People reading this paper closely will realize how much theoretical working was needed, yet hardly any  
useful results were found at the experimentally observable level. They will ask, ``what did you learn from your 
pain?". One author, who did most of the work, probably asks himself the same question. One of the other 
authors will reply with a typical German answer: ``You learn so much about NP models and field theory.'' 
\footnote{The names of the authors will not reveal to you who said it.}
The ``long suffering'' author thought of Robert Herrick, a disciple of Ben Jonson:
\begin{center}
{\bf NO PAINS, NO GAINS}\\
If little labour, little are our gains:\\
Man's fortunes are according to his pains.\\
\hspace{2.7cm}-- {\it Hesperides} 752, (1648)
\end{center}

\section{Acknowledgements}
\label{ACK}

This work was supported by the NSF under the Grant No. PHY-0807959.

\section*{Appendix A: QCD Corrections}
\addcontentsline{toc}{section}{Appendix A: QCD Corrections}
\label{APPA} 
 \renewcommand{\theequation}{A.\arabic{equation}}
  \setcounter{equation}{0}  

Here we include some of the numbers and functions which appear in the QCD correction of the operators. The function $f(z)$ in Eq.\ref{eq:fz} is that which appears in the 2-loop $O(\alpha_s)$ correction to $C_7$ is given by \cite{Greub}. 
\begin{eqnarray}
\nonumber  f(z)&=&-\frac{1}{243}\Big[576\pi ^2z^{\frac{3}{2}}+\Big(3672-288\pi ^2-1296 \zeta(3)+\left(1944-324\pi ^2\right)\log(z)\\
\nonumber&&+108\log(z)^2+36\log(z)^3\Big)z+\Big(324-576\pi ^2+\left(1728-216\pi ^2\right)\log(z)\\
\nonumber&&+324\log(z)^2+36\log(z)^3\Big)z^2+\Big(1296-12\pi ^2+1776\log(z)-2052\log(z)^2\Big)z^3\Big]\\
\nonumber&&-\frac{4\pi  i}{81}\Big[\Big(144-6\pi ^2+18\log(z)+18\log(z)^2\Big)z+\Big(-54-6\pi ^2+108\log(z)\\
&&+18\log(z)^2\Big)z^2+\Big(116-96\log(z)\Big)z^3\Big]+O(z^4)\label{eq:fz}
\end{eqnarray}
The vectors ${\bf a}$ and ${\bf z}$ \cite{Greub} which appear in $C_7$ are given by Eq.\ref{eq:Wil16} and Eq.\ref{eq:Wil17}. The matrix {\bf X} is given by Eq.\ref{eq:Wil18}.
\begin{eqnarray}
{\bf a}&=&\left\{\frac{14}{23},\frac{16}{23},\frac{6}{23},-\frac{12}{23},0.4086,-0.4230,-0.8994,0.1456\right\}\label{eq:Wil16}\\
{\bf z}&=&\left\{\frac{14}{25},\frac{16}{25},\frac{6}{25},-\frac{12}{25},0.3469,-0.4201,-0.8451,0.1317\right\}\label{eq:Wil17}\\
{\bf X}&=&\left(
\begin{array}{cccccccc}
 -3.5687 & 2.5813 & 0.4 & 0. & 0.6524 & -0.0532 & -0.0034 & -0.0084 \\
 -4.0742 & 2.7827 & 0.4 & 0. & 0.8461 & 0.0444 & 0.0068 & -0.0059 \\
 -22.423 & 18.290 & 0. & 0. & 4.3019 & -0.1241 & 0.0001 & -0.0452 \\
 -23.434 & 18.693 & 0. & 0. & 4.6894 & 0.071 & 0.0206 & -0.0402 \\
 9.8081 & -8.8366 & 0. & 0. & -0.7779 & 0.0289 & -0.0486 & -0.1739 \\
 2.8271 & -3.2361 & 0. & 0. & 0.4903 & 0.0433 & -0.1303 & 0.0056 \\
\end{array}
\right)\label{eq:Wil18}
\end{eqnarray}
 The Wilson coefficients of the operators $O_{1-6}$ at $\mu=m_b$ are given by \cite{BurasTU}
\begin{eqnarray}
C_j(m_b)=\sum_{i=3}^8h_{j(i-2)}\eta_b^{a_i} \phantom{xxx}\forall j=1\ldots 6\label{eq:WC}
\end{eqnarray}
with the initial condition
\begin{eqnarray}
{\bf C}(m_W)=\{0,1,0,0,0,0\}
\end{eqnarray}
The coefficient matrix $h_{ji}$ given by
 \begin{eqnarray}
\left( \begin{array}{cccccc}
1/2 &  -1/2 & 0 & 0 & 0& 0\\
1/2 &  1/2 & 0 & 0 & 0& 0\\
1/6 & -1/14 & 0.0510 & -0.1403 & -0.0113 & 0.0054\\
-1/6 & -1/14 & 0.0984 & 0.1214 & 0.0156 & 0.0026\\
0 & 0 & -0.0397 & 0.0117 & -0.0025 & 0.0304\\
0 & 0 & 0.0335 & 0.0239 & -0.0462 & -0.0112\\
 \end{array}\right)
 \end{eqnarray}

The Wilson coefficients given in Eq.\ref{eq:WC} can be calculated using the anomalous dimension matrix given in Ref.\cite{Greub} with five active flavors for $\mu\ge m_b$. To calculate ${\bf C}(m_c)$ the same procedure with four active flavors for $m_b\ge\mu\ge m_c$ will yeild the necessary results. 

\fancyhf{}
 
\lhead{\uppercase{Appendix}}
\cfoot{\thepage}
\section*{Appendix B: Auxiliary Functions}
\label{APPB}
\addcontentsline{toc}{section}{Appendix B: Auxiliary Functions}
 \renewcommand{\theequation}{B.\arabic{equation}}
  \setcounter{equation}{0}  
  
The auxiliary functions that come into the LHT contributions to $D^0\to X_u l^+l^-$ are listed below. They can also be found in \cite{rareBuras}.

\begin{tabular}{ll}
$G_Z(x,y)= -\frac{3}{4}U(x,y,1)$&$G_A(x,y)=\frac{1}{25a}G_Z(x,y)$\\
&\\
$G_\eta(x,y)=-\frac{3}{10a}U(x,y,\eta_a)$&$F_W(x,y)=\frac{3}{2}x-F_5(x,y)+7F_6(x,y)+3U(x,y,1)$\\
&\\
$F_2(x)=-\frac{1}{2}\left(\frac{x^2\log(x)}{(1-x)^2}+\frac{1}{1-x}\right)$&$F_5(x,y)=\frac{x^3\log(x)}{(1-x)(y-x)}-\frac{y^3\log(y)}{(1-y)(x-y)}$\\
&\\
$F_6(x,y)=\frac{z^2\log(x)}{(1-x)(y-x)}+\frac{y^2\log(y)}{(1-y)(x-y)}$&$R_2(x)=-\left(\frac{x \log(x)}{(1-x)^2}+\frac{1}{1-x}\right)$\\
&\\
$S(x)=x\left(\frac{x^2-2x+4}{(1-x)^2}\log(z)+\frac{7-x}{2(1-x)}\right)$&$U(x,y,\eta)=\frac{x^2\log(x)}{(1-x)(\eta-x)(x-y)}+\frac{y^2\log(y)}{(1-y)(\eta-y)(y-x)}+\frac{\eta^2\log(\eta)}{(1-\eta)(x-\eta)(\eta-y)}$
\end{tabular}
\begin{equation}
\end{equation}


The function $S(x)$ is a contribution from the $Z_L$ penguin diagrams with internal mirror quarks which was pointed out in \cite{Goto} and subsequently in \cite{AgI}. It replaces the divergence mentioned in \cite{rareBuras} which was later updated in \cite{BlankeUpdate}.

\fancyhf{}
 
\lhead{\uppercase{References}}
\cfoot{\thepage}

\end{document}